\documentclass[journal,doublecolumn]{IEEEtran}
\usepackage{graphicx}
\usepackage{subfig}
\usepackage{amsmath}
\usepackage{amsthm} 
\usepackage{amssymb}
\usepackage[font=footnotesize]{caption} 
\usepackage{subfig} 
\usepackage[noadjust]{cite} 
\usepackage{color}
\usepackage{algorithm} 
\usepackage{algpseudocode} 
\usepackage{enumerate}
\usepackage[hidelinks,colorlinks=false]{hyperref}
\usepackage{authblk} 

\usepackage{bm}
\usepackage{tikz}
\usetikzlibrary{calc} 
\usetikzlibrary{shapes} 
\usetikzlibrary{chains}
\usetikzlibrary{fit}
\usetikzlibrary{arrows}
\usetikzlibrary{decorations.text} 
\usetikzlibrary{decorations.markings}
\usetikzlibrary{decorations.pathmorphing} 
\usetikzlibrary{shadows}
\usetikzlibrary{patterns}
\usetikzlibrary{matrix}
\usepackage{pgfplots}
\usepackage[europeanresistors]{circuitikz}
\usepackage[outline]{contour} 
\contourlength{1.5pt}

\newtheorem{definition}{Definition}

\newcommand{\Null}{\mathrm{Null}}
\newcommand{\Range}{\mathrm{Range}}
\newcommand{\one}{\mathbf{1}}
\newcommand{\rank}{\mathrm{rank}}
\newcommand{\myspan}{\mathrm{span}}
\newcommand{\mydiag}{\mathrm{diag}}
\newcommand{\diff}{\mathrm{d}}

\newcommand{\R}{\mathbb{R}}
\newcommand{\G}{\mathcal{G}}
\newcommand{\E}{\mathcal{E}}
\newcommand{\V}{\mathcal{V}}
\newcommand{\N}{\mathcal{N}}

\renewcommand{\L}{\mathcal{B}}

\newcommand{\sk}[1]{\left[#1\right]_\times} 

\graphicspath{{figures/}}

\title{Bearing Rigidity Theory and its Applications for Control and Estimation of Network Systems\\
{\Large Life Beyond Distance Rigidity}
}
\author{Shiyu Zhao \qquad Daniel Zelazo
}

\begin{document}

\maketitle
Distributed control and estimation of multi-agent systems has received tremendous research attention in recent years due to their potential across many application domains \cite{Cao2013OverviewMultiagent,Oh2015Automatica}. Here, the term ``agent" can represent a sensor, an autonomous vehicle, or any general dynamical system.  These multi-agent systems are becoming increasingly attractive because of their robustness against system failure, their ability to adapt to dynamic and uncertain environments, and their economic advantages compared to the implementation of more expensive monolithic systems.

Formation control and network localization are two fundamental tasks for multi-agent systems that enable them to perform complex missions. The goal of formation control is to control each agent using local information from neighboring agents so that the entire team forms a desired spatial geometric pattern (see \cite{Oh2015Automatica} for a recent survey on formation control). While the notion of a formation as a geometric pattern has a natural meaning for robotic systems, it may also correspond to more abstract configurations for the system state of a team of agents. The goal of network localization is to estimate the location of each agent in a network using locally sensed or communicated information from neighboring agents \cite{Akyildiz2002,AspnesTMC2006,Mao2007NetworkLocalization,Barooah2007}. Network localization is usually the first step that must be completed before a sensor network provides other services like positioning mobile robots or monitoring areas of interest.

For a formation control or network localization task, the type of information available to each agent is an important factor that determines the design of the corresponding control or estimation algorithms. Most of the existing approaches for formation control assume that each agent can obtain the relative positions of their nearest neighbors. In order to obtain relative positions in practice, each agent can measure their absolute positions using, for example, GPS, and then share their positions with their neighbors via wireless communications. This method is, however, not applicable when operating in GPS-denied environments such as indoors, underwater, or in deep space. Furthermore, the absolute accuracy of the GPS may not meet the requirements of high-accuracy formation control tasks. Rather than relying on external positioning systems such as GPS, each agent can use onboard sensors to sense their neighbors.

Optical cameras are widely used onboard sensors for ground and aerial vehicles to achieve various sensing tasks due to their characteristics of being low-cost, light-weight, and low-power. It is notable that optical cameras are inherently bearing-only sensors. Specifically, once a target has been recognized in an image, its bearing relative to the camera can be calculated immediately from its pixel coordinate based on the pin-hole camera model \cite[Section~3.3]{MayiBook}. As a comparison, the range from the target to the camera is more complicated to obtain because it requires additional geometric information of the target and extra estimation algorithms, which may significantly increase the complexity of the vision sensing system. Although stereo cameras can be used to estimate the range of a target by triangulating the bearings of the target \cite{Hartley1997Triangulation}, the estimation accuracy degenerates rapidly as the range of the target increases due to the short baseline between the two cameras. In summary, since it is easy for vision to measure bearings, but relatively difficult to obtain accurate range information, vision can be effectively modeled as a bearing-only sensing approach in multi-agent formation control \cite{nima2009TR,Tron2016CSM}. In addition to cameras, other types of sensors such as passive radars, passive sonars, and sensor arrays are also able to measure relative bearings \cite{BearingTracking1999,Mao2007NetworkLocalization,Liuwei2017}.

When each agent is only able to access the relative bearings to their neighbors, two types of strategies can be adopted to utilize these bearings to achieve formation control or network localization. The first strategy is to use bearings to estimate relative positions. This strategy leads to coupled control and estimation problems whose global stability is difficult to prove (see, for example, \cite{Stacey2016TAC}). Moreover, the estimation of relative positions depends on an observability condition requiring that the relative motion of each pair of neighboring agents satisfy certain conditions \cite{BearingObservability1993}. Although this observability condition can be achieved in certain applications, such as bearing-only circumnavigation \cite{Anderson2014TAC,Dimos2014US,ZhengRonghao2015Automatica,Mengbin2017TAES}, it is difficult to satisfy in general formation control tasks where all the agents are supposed to form a target formation with no relative motions among the agents. This observability condition is not satisfied either in network localization because all the sensors are stationary. The second strategy, which is the focus of this article, is to directly apply bearings in formation control or network localization without estimating relative positions. This strategy does not require relative position estimation, but it requires designing new control and estimation algorithms that only utilize bearing measurements.

The purpose of this article is to provide a tutorial overview of recent advances in the area of bearing-based formation control and network localization. The first problem addressed in this article is to understand when the formation control or network localization problems can be solved using only inter-neighbor bearing measurements. In fact, any distributed control or estimation task requires certain fundamental architectural conditions of the multi-agent system. For example, in consensus problems, a network must possess a spanning tree in order to ensure the states of different agents converge to the same value \cite{OlfatiTAC2004,LinZhiyun2005TAC,Ren2007CSM,LiZhongkui2010}.
For bearing-based formation control and network localization, there is also an architectural requirement to solve these problems - this property is known as \emph{bearing rigidity}. The bearing rigidity theory, also called parallel rigidity theory in the literature, was originally introduced for computer-aided design \cite{Whiteley1999ParallelDraw} and has received increasing attention in recent years due to its important applications in bearing-based control and estimation problems \cite{bishopconf2011rigid,Eren2012IJC,zelazo2014SE2Rigidity,TronRigidComponent,zhao2014TACBearing}. The bearing rigidity theory studies the fundamental problem of under what conditions can the geometric pattern of a network be uniquely determined if the bearing of each edge in the network is fixed.

The bearing rigidity theory can be interpreted as an analogous theory for the classic rigidity theory based on inter-neighbor distances, which is referred to as \emph{distance rigidity} theory in this article. The classic distance rigidity theory studies the problem of under what conditions can the geometric pattern of a network be uniquely determined if the length (distance) of each edge in the network is fixed. It is a combinatorial theory for characterizing the stiffness or flexibility of structures formed by rigid bodies connected by flexible linkages or hinges.
The study of distance rigidity has a long history as a formal mathematical discipline \cite{Asimow1978Rigidty,ConnellyBook,Hendrickson1992SIAM,Connelly2005Generic,Jacobs1997,WhiteleyHenneberg1985,Whiteley2005Pseudotriangulation,Jackson2007NotesRigidity}. In recent years, it has played a fundamental role in distance-based formation control \cite{Anderson2008CSM,Krick2009IJC,oh2013IJRNC,Tian2013Global,zelazo2015Rigidity,Mou2016TAC,ChenXudong2016SIAM,ZhiyongAutomatica2016,Marina2016TR} and distance-based network localization \cite{Eren2004,AspnesTMC2006,Mao2007NetworkLocalization}. One goal of this article is to compare the distance and bearing rigidity theories by highlighting their similarities and differences.

This article addresses three important applications of the bearing rigidity theory in the area of the distributed control and estimation of multi-agent systems, briefly described below.
\begin{enumerate}[(a)]
\item \emph{Bearing-Based Network Localization}:
Consider a network of stationary nodes where only a subset of the nodes know their own absolute positions - these special nodes are referred to as the \emph{anchors} while the other nodes are called \emph{followers}.
Suppose each follower node is able to measure the relative bearings of their neighbors and share the estimates of their own positions with their neighbors by wireless communication. The aim of bearing-based network localization is to localize the follower nodes using the bearing measurements and the anchors' absolute positions \cite{eren2003,Barooah2007,BishopTAES2009,Piovan2013Automatica,Shames2013TAC,ZhuGuangwei2014Automatica,Lin2016TSP}.
Here the network localization problem may also be called network self-localization, which is usually the first step for a sensor network to provide other services such as positioning or monitoring. Network localization is essential for sensor networks in environments where GPS signals are not available, reliable, or sufficiently accurate.

\item \emph{Bearing-Based Formation Control}:
Consider a group of mobile agents where each agent is able to obtain the relative positions of their neighbors. The aim of bearing-based formation control is to steer the agents from some initial spatial configuration to a target formation with a desired geometric pattern predefined by inter-neighbor bearings \cite{bishopconf2011rigid,Bishop2015Relaxed,zhao2015ECC,zhao2015Maneuver,Fathian2016ACC}. Since the target formation is invariant to scaling variations, bearing-based formation control provides a simple solution for formation scale control, which is a practically useful technique to adjust the scale of a formation so that the agents can dynamically respond to the environment to achieve, for example, obstacle avoidance such as passing through narrow passages \cite{Han2016Scale,Coogan2012Scale}.
Note that the bearing-based formation control problem is dual to the bearing-based network localization problem. When the agent dynamics are modeled as single integrators and the leaders are stationary, the two problems are indeed identical. However, this article also considers a broader range of cases in the formation control problem - namely formation maneuvering using leaders, and different models for the agent dynamics, including double integrators and unicycles.

\item \emph{Bearing-Only Formation Control}:
The aim of bearing-only formation control is to steer a group of mobile agents to form a desired geometric pattern predefined by inter-neighbor bearings. Unlike bearing-based formation control, bearing-only formation control only requires each agent to measure the relative bearings of their neighbors, whereas relative positions are not required to be measured or estimated \cite{bishop2010SCL,Eren2012IJC,Franchi2012IJRR,Cornejo2013IJRR,zhao2013SCLDistribued,zhao2013IJCFinite,Eric2014ACC,Tron2016CSM,Tron2016CDC}. Bearing-only formation control provides a novel framework for implementing vision-based formation control tasks where vision may be modeled as a bearing-only sensing approach. It also suggests that distance information may be redundant to achieve certain formation control tasks.
\end{enumerate}
The notations for networks and formations used throughout this article are given in ``\nameref{sidebar_notationsforNetworksFormations}''.

\section*{\textbf{Bearing Rigidity Theory}}\label{section_bearingRigidityTheory}

\begin{figure}
  \centering
\includegraphics[width=\linewidth]{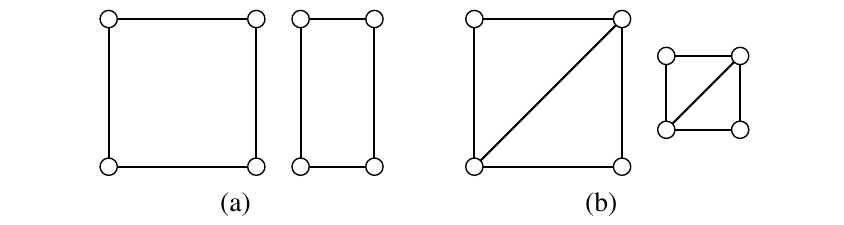}
  \caption{Illustration of bearing rigidity. The networks in (a) are not bearing rigid because the same inter-neighbor bearings may lead to different geometric patterns of the networks, for example, a square on the left and a rectangle on the right. The networks in (b) are bearing rigid because the same inter-neighbor bearings imply the same geometric pattern though the networks may differ in terms of translation and scale.}
  \label{fig_demoRigidityConcept}
\end{figure}

The bearing rigidity theory studies the problem of under what conditions the geometric pattern of a network can be uniquely determined if the bearing of each edge in the network is fixed. Equivalently stated, bearing rigidity studies as under what conditions do two networks have the same geometric pattern if they have the same bearings. To illustrate this idea, the two networks in Figure~\ref{fig_demoRigidityConcept}(a) have the same bearings but different geometric patterns. As a result, they are not bearing rigid. The two networks in Figure~\ref{fig_demoRigidityConcept}(b) have the same bearings and the same geometric pattern (modulo a scaling and a translational factor). The two networks can be shown to be bearing rigid and the rigorous proof of this result relies on the theory presented in this section.

There are three different notions of bearing rigidity: bearing rigidity, global bearing rigidity, and infinitesimal bearing rigidity. The first two are not of practical interest because they cannot ensure unique geometric patterns of networks. The third, infinitesimal bearing rigidity, is the most important one whose properties are discussed in detail in this section.
The precise definitions of the three types of bearing rigidity are given in ``\nameref{sidebar_bearingRigidity}.'' These definitions are analogous to those in the distance rigidity theory, which are listed in ``\nameref{sidebar_distanceRigidity}'' for the purpose of comparison. It is worth noting that an orthogonal projection matrix plays a key role in the bearing rigidity theory. The properties of the projection matrix are summarized in ``\nameref{sidebar_projectionOperator}.''
Moreover, note that a bearing, which is represented by a unit vector, must be expressed in a specific reference frame. In this article, the bearings in a network are all expressed in a common reference frame.

\subsection*{\textbf{Properties of Infinitesimal Bearing Rigidity}}

\begin{figure}
  \centering
\includegraphics[width=\linewidth]{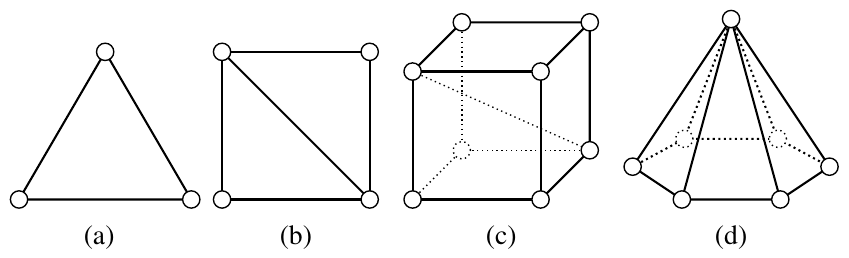}
  \caption{
  Examples of infinitesimally bearing rigid networks.
  The networks in (a) and (b) are two-dimensional and the networks in (c) and (d) are three-dimensional. It can be verified that each of these networks satisfies $\rank(\L)=dn-d-1$. The networks in (a), (b), and (c) also satisfy the Laman condition and can therefore be generated using a Henneberg construction.  Note that the two networks in (c) and (d) are infinitesimal bearing rigid but not infinitesimal distance rigid.}
  \label{fig_IPRExamples}
\end{figure}

Infinitesimal bearing rigidity has two key properties. The first is a geometric property \cite[Theorem~6]{zhao2014TACBearing} that the positions of the nodes in a network can be uniquely determined up to a translational and scaling factor by the bearings if and only if the network is infinitesimally bearing rigid.
The second is an algebraic property \cite[Theorem~4]{zhao2014TACBearing} that a network is infinitesimally bearing rigid in $d$-dimensional space if and only if the bearing rigidity matrix $R_B$ satisfies
\begin{align}\label{eq_nullConditionIBR}
\Null(R_B)=\myspan\{\one_n\otimes I_d, p\},
\end{align}
or equivalently,
\begin{align}\label{eq_rankConditionIBR}
\rank(R_B)=dn-d-1.
\end{align}
The definition of the bearing rigidity matrix $R_B$ is given in ``\nameref{sidebar_bearingRigidity}.''
Due to the above two properties, infinitesimal bearing rigidity not only ensures the unique geometric pattern of a network, but also can be conveniently examined by a mathematical condition. Examples of infinitesimally bearing rigid networks are given in Figure~\ref{fig_IPRExamples}.

The notion of infinitesimal bearing rigidity is defined based on the bearing rigidity matrix. The term ``infinitesimal'' is due to the fact that the bearing rigidity matrix is the first-order derivative (the Jacobian) of the bearing vectors with respect to the positions of the nodes.
It must be noted that infinitesimal bearing rigidity is a \emph{global} property in the sense that the bearings can \emph{uniquely} determine the geometric pattern of a network. The term ``infinitesimal'' may be dropped in this article when the context is clear.

\begin{figure}
  \centering
  \includegraphics[width=\linewidth]{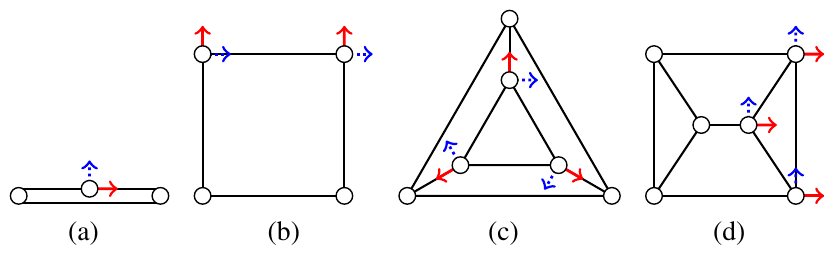}
  \caption{Examples of non-infinitesimally bearing rigid networks. The red and solid arrows represent nontrivial infinitesimal bearing motions that preserves all the inter-neighbor bearings.
  These networks are not infinitesimally distance rigid either because they have nontrivial infinitesimal distance motions (see the blue/dotted arrows).
  Note that the infinitesimal distance motions are perpendicular to the infinitesimal bearing motions.}
  \label{fig_nonIBRExamples}
\end{figure}

An infinitesimal bearing motion of a network is a motion of some nodes that preserves all the bearings. All the infinitesimal bearing motions of a network form the null space of the bearing rigidity matrix. There are two types of trivial infinitesimal bearing motions: translational and scaling motions of the entire network. These two types of trivial motions corresponds to the vectors in $\myspan\{\one_n\otimes I_d,p\}$. As a result, the rank condition in \eqref{eq_nullConditionIBR} means that a network is infinitesimally bearing rigid if and only if all infinitesimal bearing motions are trivial. This provides an intuitive way to examine bearing rigidity. For example, the networks in Figure~\ref{fig_nonIBRExamples} are not bearing rigid because they have non-trivial infinitesimal bearing motions.

An alternative necessary and sufficient condition for infinitesimal bearing rigidity is based on a special matrix termed the \emph{bearing Laplacian} \cite{zhao2015NetLocalization}. The bearing Laplacian of a network can be viewed as a weighted graph Laplacian matrix with weights that are matrices \cite{BookGodsilGraph}; thus the bearing Laplacian not only describes the topological structure of the network, but also the values of the edge bearings. The definition and properties of bearing Laplacian are summarized in ``\nameref{sidebar_bearingLaplacian}.''
For a network with an undirected graph, the bearing Laplacian has the same rank and null space as the bearing rigidity matrix \cite[Lemma~2]{zhao2015NetLocalization}. It then follows from \eqref{eq_nullConditionIBR} and \eqref{eq_rankConditionIBR} that a network is infinitesimally bearing rigid if and only if
\begin{align*}
\Null(\L)=\myspan\{\one_n\otimes I_d, p\},
\end{align*}
or equivalently,
\begin{align*}
\rank(\L)=dn-d-1.
\end{align*}
Compared to the bearing rigidity matrix, the bearing Laplacian is more convenient to use because it is symmetric and positive semi-definite for undirected graphs. When the underlying graph is directed, the bearing Laplacian and the bearing rigidity matrix may have different ranks and null spaces \cite[Theorem~4]{zhao2015CDC}.

\subsection*{\textbf{Construction of Infinitesimally Bearing Rigid Networks}}

The previous discussion provided an overview of the properties defining a bearing rigid network.  It is also of interest to explore how to construct a bearing rigid network by adding well-placed edges and nodes in a network.
Although a network is jointly characterized by its underlying graph and the configuration of the nodes, the infinitesimal bearing rigidity of a network is primarily determined by the underlying graph rather than its configuration \cite[Lemma~2]{zhao2017CDCLaman}. Given a graph, if there exists at least one configuration such that the network is infinitesimally bearing rigid, then for almost all configurations the corresponding networks are infinitesimally bearing rigid. Such graphs are called \emph{generically bearing rigid} \cite{zhao2017CDCLaman}. If a graph is not generically bearing rigid, then the corresponding network is not infinitesimal bearing rigid for any configurations. As a result, the key to construct infinitesimally bearing rigid networks is to construct generically bearing rigid graphs.

One of the most well-known methods for rigid graph construction is the Henneberg construction, originally proposed for the distance rigidity theory \cite{WhiteleyHenneberg1985}. A Henneberg construction starting from an edge connecting two vertices results in a \emph{Laman graph} \cite{Laman1970}. For a tutorial on Laman graphs and Henneberg construction, see ``\nameref{sidebar_LamanGraph}.''

\begin{figure}[t]
  \centering
\includegraphics[width=\linewidth]{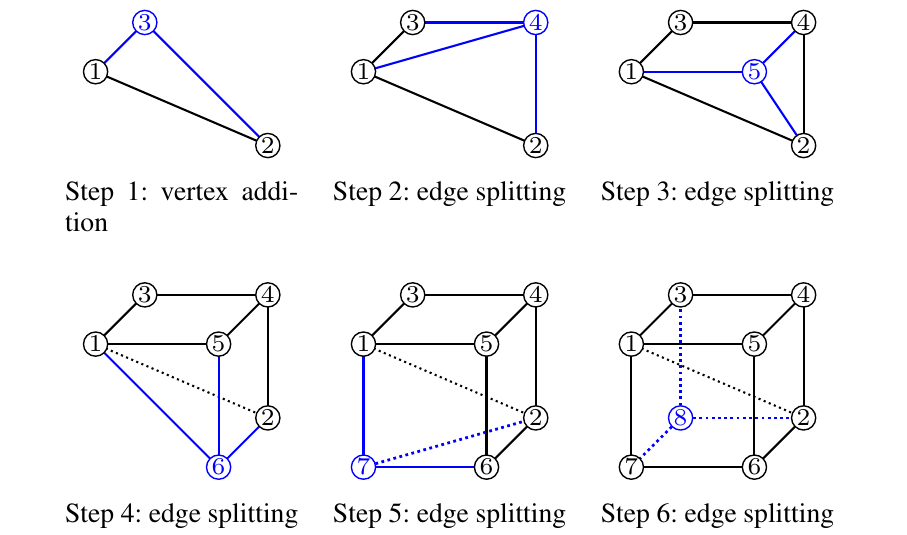}
  \caption{Illustration of the Henneberg construction procedure.  The Henneberg construction consists of two basic operations: vertex addition and edge splitting.  In this example, the procedure is used to generate an infinitesimally bearing rigid network in a three-dimensional ambient space. At each step, the underlying graph of the network is Laman.}
  \label{fig_bearingRigidNetworkConstructionEachStep}
\end{figure}

\begin{figure}[t]
  \centering
\includegraphics[width=\linewidth]{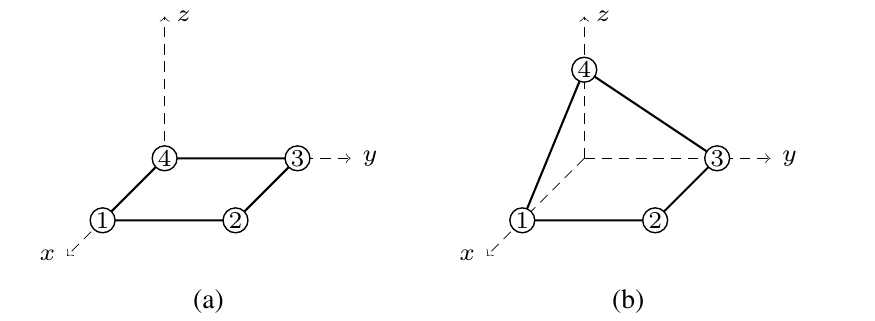}
  \caption{Example of generically bearing rigid graphs that are not Laman.  The configuration (a) is in the $x$--$y$ plane and the network is not bearing rigid. The configuration (b) is three-dimensional and the network is bearing rigid. It can be verified that $\rank(\L)=dn-d-1$ for the configuration in (b).}
  \label{fig_specialRigidExample}
\end{figure}

In the bearing rigidity theory, the main result about Laman graphs is that all Laman graphs are generically bearing rigid in arbitrary dimensions  \cite[Theorem~1]{zhao2017CDCLaman}. That means if the underlying graph of a network is Laman, then the network is infinitesimally bearing rigid for almost all configurations in an arbitrary dimension.
Figure~\ref{fig_bearingRigidNetworkConstructionEachStep} illustrates the Henneberg construction procedure for a three-dimensional infinitesimally bearing rigid network whose underlying graph is Laman. Note that the Laman condition is merely sufficient but not necessary for generic bearing rigidity. A counterexample is given in Figure~\ref{fig_specialRigidExample}, where the graph is generically bearing rigid but not Laman. However, for networks in the plane, a graph is generically bearing rigid if and only if it is Laman \cite[Theorem~2]{zhao2017CDCLaman}.

Since a Laman graph has $2n-3$ edges where $n$ is the number of nodes, $2n-3$ edges are sufficient to guarantee the bearing rigidity of a network in an arbitrary dimension. For example, every network in Figure~\ref{fig_bearingRigidNetworkConstructionEachStep} is bearing rigid in the three dimensional space and has $2n-3$ edges.
It must be noted that $2n-3$ is not the minimum number of edges required to ensure bearing rigidity. The counterexample given in Figure~\ref{fig_specialRigidExample} shows that a graph with less than $2n-3$ edges may be generically bearing rigid in three dimensions. It is still an open problem to construct all generically bearing rigid graphs up to now.
A comparison between the bearing and distance rigidity theories is given in ``\nameref{sidebar_compareBearingDistanceRigidity}.''

\section*{\textbf{Bearing-Based Network Localization}}\label{section_bearingNetworkLocalization}

This section introduces the theory of bearing-based network localization that addresses two fundamental problems.
The first problem is \emph{localizability}, which describes whether or not a network can possibly be localized. The second problem is \emph{how} to localize a network in a distributed manner if it is localizable.

Consider a network of nodes where the first $n_a$ nodes are anchors and the remaining $n_f$ ($n_f=n-n_a$) nodes are followers. Let $\V_a=\{1,\dots,n_a\}$ and $\V_f=\V\setminus \V_a$ be the sets of anchors and followers, respectively. The true positions of the leaders and followers are denoted as $p_a=[p_1^T,\dots,p_{n_a}^T]^T$ and $p_f=[p_{n-n_a}^T,\dots,p_{n}^T]^T$, respectively. The aim of network localization is to determine the positions of the followers $\{p_i\}_{i \in \V_f}$ using the edge bearings $\{g_{ij}\}_{(i,j)\in\E}$ and the positions of the anchors $\{p_i\}_{i\in\V_a}$.
All the inter-neighbor bearings are expressed in a common reference frame.

\subsection*{\textbf{Bearing-Based Localizability}}

\begin{figure}
  \centering
\includegraphics[width=\linewidth]{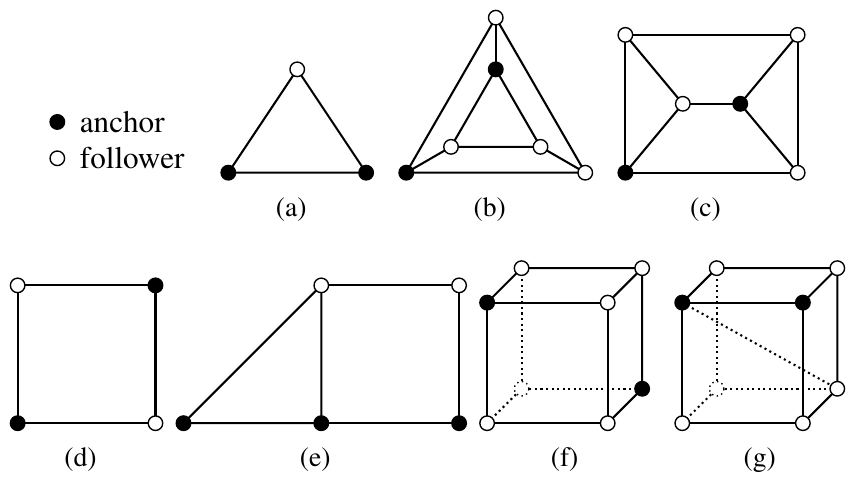}
  \caption{Examples of bearing localizable networks.
  The networks are localizable because $\L_{ff}$ of each network is nonsingular.
  The intuitive interpretation is that every infinitesimal bearing motion involves at least one anchor.
  Note that the networks in (b)-(f) are not infinitesimally bearing rigid but they are localizable.}
  \label{fig_Example_localizable}
\end{figure}

\begin{figure}
  \centering
\includegraphics[width=\linewidth]{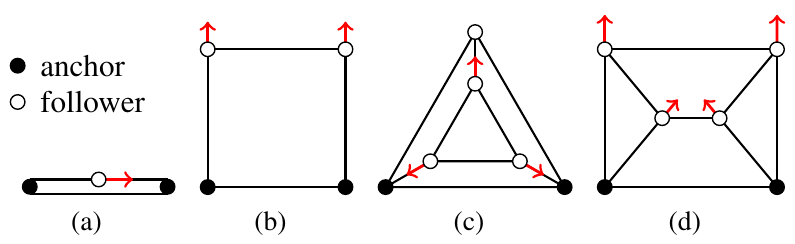}
  \caption{Examples of networks that not bearing localizable.
  The black solid dots represent the anchors and the white dots for followers.
  The networks are not localizable because $\L_{ff}$ of each network is singular.
  The intuitive interpretation is that the networks have infinitesimal bearing motions that only correspond to the followers (see the red arrows).}
  \label{fig_Example_nonlocalizable}
\end{figure}

Localizing the follower nodes is to solve for $\hat{p}_i$, the estimate of $p_i$, for all $i\in\V_f$, obtained from the set of nonlinear equations,
\begin{align}\label{eq_networkLocalization_nonlinearConstraint}
  \left\{\begin{array}{ll}
    \displaystyle{\frac{\hat{p}_j-\hat{p}_i}{\|\hat{p}_j-\hat{p}_i\|}=g_{ij},} & (i,j)\in\E, \\
    \hat{p}_i=p_i, & i\in\V_a. \\
  \end{array}\right.
\end{align}
The true location $p$ of the network is a feasible solution to \eqref{eq_networkLocalization_nonlinearConstraint}. However, there may exist an infinite number of other feasible solutions. This leads to the definition of localizability. A network $(\G,p)$ is called \emph{bearing localizable} if the true position $p$ is the unique feasible solution to \eqref{eq_networkLocalization_nonlinearConstraint}.
It can be further shown that $p$ is the unique solution to \eqref{eq_networkLocalization_nonlinearConstraint} if and only if $p$ is the unique global minimizer of the least-squares problem \cite[Lemma~1]{zhao2015NetLocalization}
\begin{align}\label{eq_optimization_noisyBearing_accuAnchor}
  \min_{\hat{p}\in\R^{dn}} & \quad J(\hat{p})=\frac{1}{2}\sum_{i\in\V}\sum_{j\in\N_i}\|P_{{g}_{ij}}(\hat{p}_i-\hat{p}_j)\|^2=\hat{p}^T \L \hat{p},
\end{align}
subject to $\hat{p}_i=p_i$ for $i\in\V_a$.
It has been proven that $p$ is the unique minimizer of \eqref{eq_optimization_noisyBearing_accuAnchor} if and only if the matrix $\L_{ff}$ is nonsingular \cite[Theorem~1]{zhao2015NetLocalization}. The definition of $\L_{ff}$ is given in ``\nameref{sidebar_bearingLaplacian}.'' When $\L_{ff}$ is nonsingular, the positions of the followers can be solved as $\hat{p}_f^*=-\L_{ff}^{-1}\L_{fa}p_a$.
Examples of bearing localizable and non-localizable networks are given in Figures~\ref{fig_Example_localizable} and \ref{fig_Example_nonlocalizable}, respectively.


While the nonsingularity of $\L_{ff}$ is an algebraic condition for bearing localizability, it does not give any intuition on what a bearing localizable network looks like. The following conditions can provide more intuition for bearing localizable networks. First of all, a necessary and sufficient rigidity condition for bearing localizability is that every infinitesimal bearing motion of a network must involve at least one anchor \cite[Theorem~2]{zhao2015NetLocalization}.
More specifically, if there exists a nonzero infinitesimal bearing motion for a network, there would exist different networks having exactly the same bearings as the true network.
As a result, infinitesimal bearing motions introduce ambiguities to the localization of the true network.
When the infinitesimal motion involves at least one anchor, the ambiguities can be resolved by the anchors whose positions are known, and hence the network location can be uniquely determined. This rigidity condition provides an intuitive way to examine network localizability (see, for example, Figure~\ref{fig_Example_nonlocalizable}).

The following condition indicates how many anchors are required to guarantee bearing localizability. The number of anchors in a bearing localizable network in $\R^d$ must satisfy \cite[Corollary~1]{zhao2015NetLocalization}
\begin{align}\label{eq_necessaryConditionAnchorNum}
    n_a\ge \frac{\dim\left(\Null(\L)\right)}{d}\ge \frac{d+1}{d}.
\end{align}
Inequality~\eqref{eq_necessaryConditionAnchorNum} has two important implications.
The first is that every bearing localizable network must have at least \emph{two} anchors because $(d+1)/d>1$.
The second is that more anchors are required when the ``degree of bearing rigidity'' of the network is weak.
Here, the ``degree of bearing rigidity'', characterized by $\dim(\Null(\L))$, is strongest if $\dim(\Null(\L))$ reaches the smallest value $d+1$ (when the network is infinitesimally bearing rigid) and weak if its value is greater than $d+1$.

\begin{figure}
  \centering
\includegraphics[width=\linewidth]{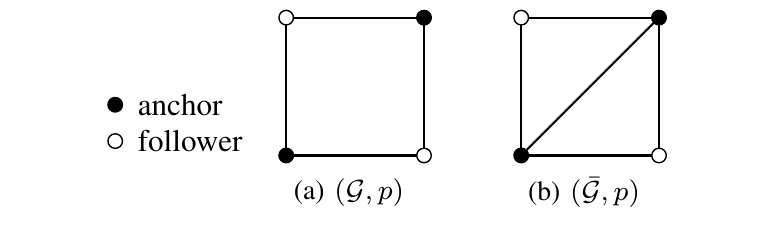}
  \caption{Illustration of an augmented network for the localization problem. The augmented network $(\bar{\G},p)$ in (b) is obtained from $(\G,p)$ by connecting each pair of anchors in $(\G,p)$. Since deleting or adding the edge between any pair of anchors only changes $\L_{aa}$ but not $\L_{ff}$, $(\G,p)$ and $(\bar{\G},p)$ have exactly the same $\L_{ff}$, and hence they have the same localizability properties.}
  \label{fig_Example_augmentedNetwork}
\end{figure}

The following two conditions explicitly address the relation between bearing localizability and bearing rigidity.
(i) A sufficient condition for a network to be bearing localizable is that it is infinitesimally bearing rigid and has at least two anchors \cite[Corollary~3]{zhao2015NetLocalization}.
The intuition behind this condition is as follows. If a network is infinitesimally bearing rigid, then it can be uniquely determined up to a translation and a scaling factor. If there are at least two anchors, the translational and scaling ambiguity can be eliminated by the anchors and thus the entire network can be fully determined.
It must be noted that infinitesimal bearing rigidity is merely sufficient but not necessary for bearing localizability. For example, the networks in Figure~\ref{fig_Example_localizable}(b)-(f) are bearing localizable but not infinitesimally bearing rigid.
(ii) Let $(\bar{\G},p)$ be the augmented network of $(\G,p)$ which is obtained from $(\G,p)$ by connecting each pair of anchors (see Figure~\ref{fig_Example_augmentedNetwork} for illustration).
Then, another sufficient condition for bearing localizability is that network $(\G,p)$ is bearing localizable if the augmented network $(\bar{\G},p)$ is infinitesimally bearing rigid and there are at least two anchors \cite[Corollary~2]{zhao2015NetLocalization}. This condition is more relaxed in the sense that it does not require $(\G,p)$ to be infinitesimally bearing rigid. When there are more than two anchors, the infinitesimal bearing rigidity of $(\bar{\G},p)$ is merely sufficient but not necessary for the bearing localizability of $(\G,p)$ (see Figure~\ref{fig_Example_localizable}(f) for a counterexample where the network is bearing localizable but the augmented network is not infinitesimally bearing rigid). When there are exactly two anchors, the infinitesimal bearing rigidity of $(\bar{\G},p)$ is both necessary and sufficient for the bearing localizability of $(\G,p)$ \cite[Theorem~3]{zhao2015NetLocalization}.

\subsection*{\textbf{Distributed Localization Protocols}}

\begin{figure}
  \centering
\includegraphics[width=\linewidth]{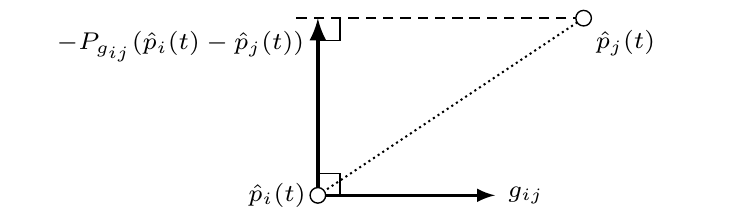}
  \caption{The geometric interpretation of the bearing-based control law in \eqref{eq_networkLocalizeProtocal}. The term $P_{g_{ij}}(\hat{p}_j-\hat{p}_i)$ is perpendicular to $g_{ij}$ and it aims to steer agent $i$ such that $\hat{g}_{ij}(t)$ aligns with $g_{ij}$.}
  \label{fig_bearingBasedLocalizationGeometricMeaning}
\end{figure}

If a network is bearing localizable, a question that follows is how to localize it in a distributed manner.
Suppose each node has an initial guess of its own position as $\hat{p}_i(0)$. The objective is to design a distributed protocol to drive $\hat{p}_i(t)\rightarrow p_i$ for all $i\in\V_f$ as $t\rightarrow\infty$.
This objective can be achieved by the protocol \cite{zhao2015NetLocalization}
\begin{align}\label{eq_networkLocalizeProtocal}
    \dot{\hat{p}}_i(t)&=-\sum_{j\in\N_i} P_{{g}_{ij}} (\hat{p}_i(t)-\hat{p}_j(t)), \quad i\in\V_f,
\end{align}
where $P_{g_{ij}}=I_d-g_{ij}g_{ij}^T$.
Protocol \eqref{eq_networkLocalizeProtocal} is actually the gradient-descent protocol for the objective function in the least-squares problem \eqref{eq_optimization_noisyBearing_accuAnchor}.
The geometric interpretation of this protocol is illustrated in Figure~\ref{fig_bearingBasedLocalizationGeometricMeaning}.
The expression of protocol \eqref{eq_networkLocalizeProtocal} is similar to the well-known linear consensus protocols \cite{OlfatiTAC2004,Ren2007CSM}. The difference is that the weight for each edge in \eqref{eq_networkLocalizeProtocal} is an orthogonal projection matrix, while in the consensus protocols, the weight for each edge is a scalar. This important distinction leads to very different properties of the dynamical system. The unique structure of the projection matrix is the key feature that enables \eqref{eq_networkLocalizeProtocal} to solve the bearing-based network localization problem.

\begin{figure}
  \centering
  \subfloat[Initial and final estimates.]{\includegraphics[width=.75\linewidth]{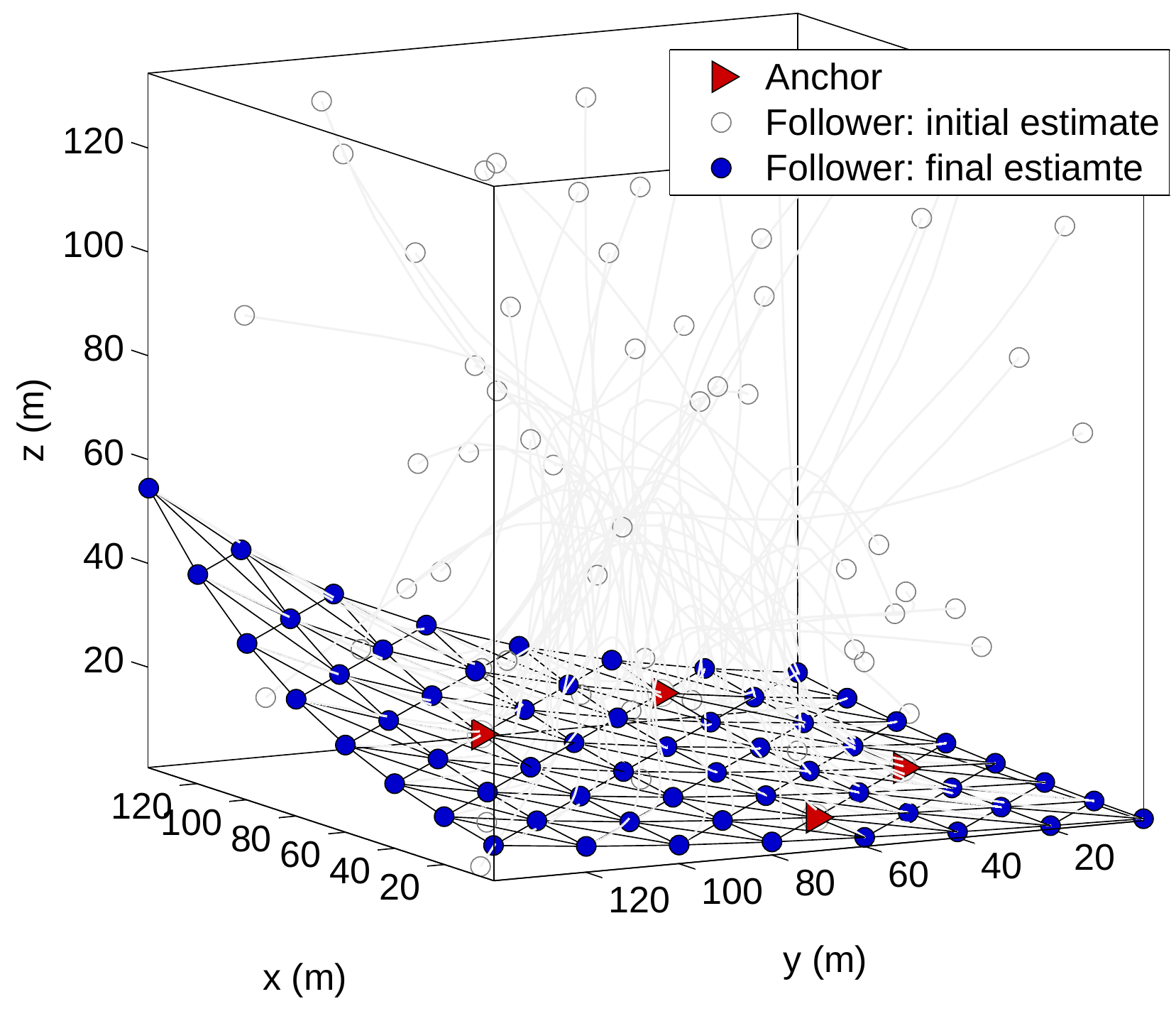}}\\
  \subfloat[Localization error $\|\hat{p}_i(t)-p_i\|$.]{\includegraphics[width=.75\linewidth]{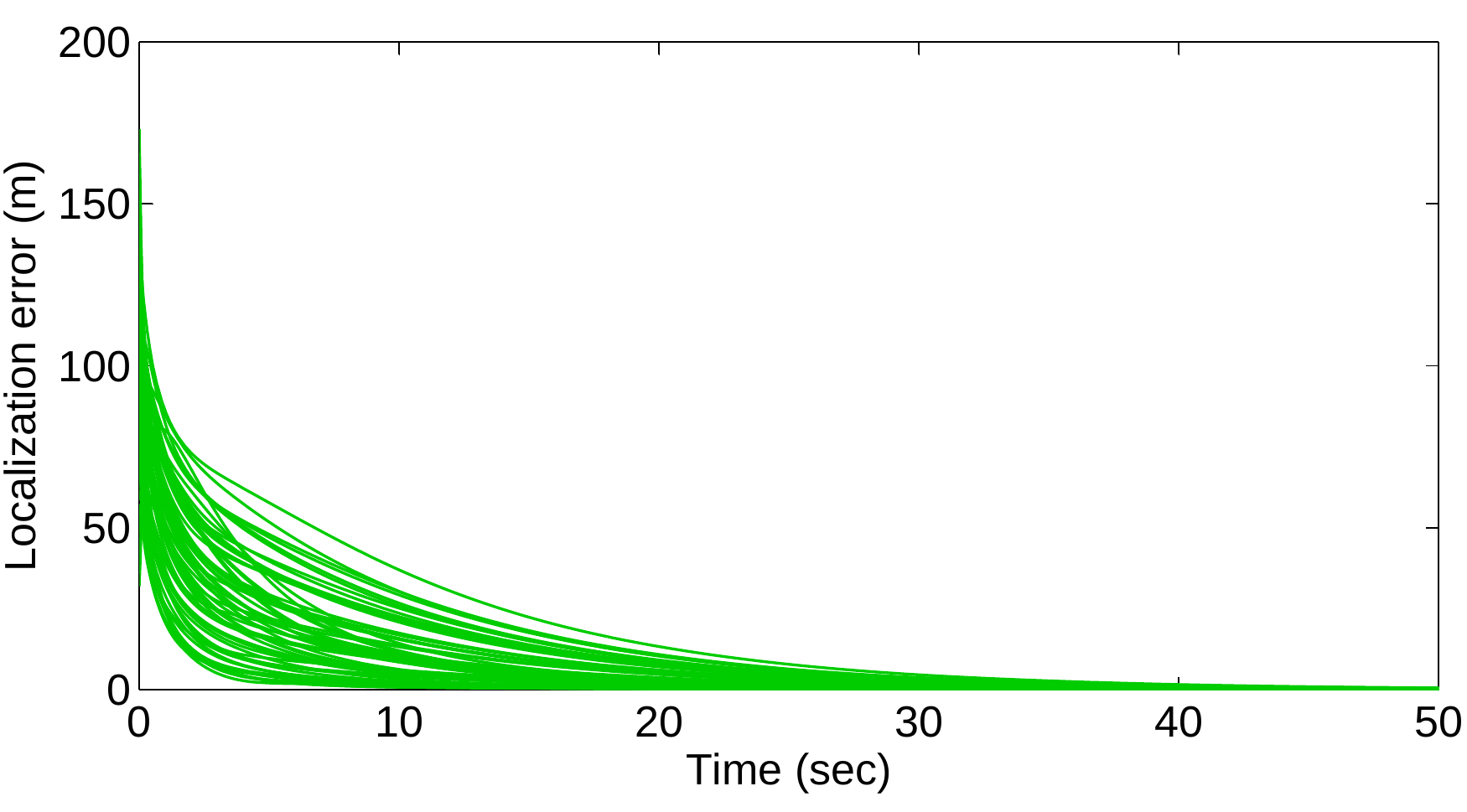}}
  \caption{Simulation example to demonstrate the localization protocol in \eqref{eq_networkLocalizeProtocal}.
  The real network is located on a three-dimensional surface. It consists of $210$ edges and $64$ nodes, four of which are anchors. The network is infinitesimally bearing rigid because $\rank(\L)=188=dn-d-1$. Therefore, the network is localizable since there are more than two anchors. As can be seen, given a random initial guess, the localization error of each node converges to zero.}
  \label{fig_sim_networkLocalization}
\end{figure}

The compact matrix form of \eqref{eq_networkLocalizeProtocal} is
$$\dot{\hat{p}}_f(t)=-\L_{ff}\hat{p}_f(t)-\L_{fa}p_a,$$
where $\L$ is the bearing Laplacian of the true network.
This protocol can globally localize the network if and only if the network is bearing localizable (that is $\L_{ff}$ is nonsingular) \cite[Theorem~4]{zhao2015NetLocalization}.
Figure~\ref{fig_sim_networkLocalization} shows a simulation example to demonstrate protocol \eqref{eq_networkLocalizeProtocal}.
The impact of measurement noise on bearing-based network localization has been discussed in \cite{zhao2015NetLocalization}.

\section*{\textbf{Bearing-Based Formation Control}}\label{section_bearingBasedFormationControl}

This section introduces the theory of bearing-based formation control, which studies how to steer a group of agents to achieve a bearing-constrained target formation using relative position measurements.
In particular, consider a group of mobile agents where the first $n_\ell$ agents are leaders and the remaining $n_f$ ($n_f=n-n_\ell$) agents are followers. Let $\V_\ell=\{1,\dots,n_\ell\}$ and $\V_f=\V\setminus \V_\ell$ be the sets of leaders and followers, respectively. The positions of the leaders and followers are denoted as $p_\ell=[p_1^T,\dots,p_{n_\ell}^T]^T$ and $p_f=[p_{n-n_\ell}^T,\dots,p_{n}^T]^T$, respectively.
The target formation is specified by the constant bearing constraints $\{g_{ij}^*\}_{(i,j)\in\E}$ and the leader positions $\{p_i(t)\}_{i\in\V_\ell}$. The control objective is to control the positions of the followers $\{p_i(t)\}_{i \in \V_f}$ such that $g_{ij}(t)\rightarrow g_{ij}^*$ as $t\rightarrow \infty$ for all ${(i,j)\in\E}$. All the bearings are expressed in a common reference frame.

\subsection*{\textbf{Bearing-Based Formation Control of Single Integrators}}

First, consider the case where the dynamics of each mobile agent can be modeled as the single integrator
\begin{align*}
\dot{p}_i(t)=u_i(t),
\end{align*}
where $u_i(t)$ is the velocity input to be designed.
If the leaders are stationary, the bearing-based formation control problem can be solved by \cite{zhao2015ECC}
\begin{align}\label{eq_bearingBasedControlLaw_singleIntegrator_stabilization}
    \dot{p}_i(t)=-\sum_{j\in\N_i} P_{g_{ij}^*} (p_i(t)-p_j(t)),\quad i\in\V_f,
\end{align}
where $P_{g_{ij}^*}=I_d-g_{ij}^*(g_{ij}^*)^T $.
The matrix form of the control law is
$$\dot{p}_f(t)=-\L_{ff}p_f(t)-\L_{f\ell}p_\ell,$$
where $\L$ is the bearing Laplacian of the target formation.
Control law \eqref{eq_bearingBasedControlLaw_singleIntegrator_stabilization} can globally stabilize a target formation if and only if the target formation is bearing localizable (that is $\L_{ff}$ is nonsingular) \cite{zhao2015ECC}.
Note that control law \eqref{eq_bearingBasedControlLaw_singleIntegrator_stabilization} has a similar expression to the network localization protocol in \eqref{eq_networkLocalizeProtocal}.
In fact, the bearing-based formation control problem is mathematically equivalent to the bearing-based network localization problem when the target formation is stationary and each agent is a single integrator.

If the leaders move at a constant nonzero speed, control law \eqref{eq_bearingBasedControlLaw_singleIntegrator_stabilization} would yield a constant nonzero tracking error. The tracking error may be eliminated by using the following proportional-integral control law proposed in \cite{zhao2015MSC},
\begin{align}\label{eq_bearingBasedControlLaw_singleIntegrator_maneuverPI}
    \dot{p}_i(t)
    &=- \sum_{j\in\N_i} P_{g_{ij}^*} \Big[k_p(p_i(t)-p_j(t)) \Big.\nonumber \\
    &\qquad\qquad \Big.-k_I\int_0^t (p_i(\tau)-p_j(\tau))\diff \tau\Big], \quad i\in\V_f,
\end{align}
where $k_p$ and $k_I$ are constant positive control gains. The target formation is globally stable under the action of control law \eqref{eq_bearingBasedControlLaw_singleIntegrator_maneuverPI} if and only if it is bearing localizable \cite{zhao2015MSC}.

If the leader velocities are time-varying, control law \eqref{eq_bearingBasedControlLaw_singleIntegrator_maneuverPI} would fail to ensure zero tracking errors. The time-varying case can be handled by the following control law that requires velocity feedback:
\begin{align}\label{eq_bearingBasedControlLaw_singleIntegrator_maneuverPD}
    \dot{p}_i(t)
    =- K_i^{-1}\sum_{j\in\N_i} P_{g_{ij}^*} \left[k_p(p_i(t)-p_j(t))-\dot{p}_j(t)\right], \; i\in\V_f,
\end{align}
where $K_i=\sum_{j\in\N_i}P_{g_{ij}^*}$.
The stability of control law \eqref{eq_bearingBasedControlLaw_singleIntegrator_maneuverPD} can be proven as below. First, the nonsingularity of $K_i$ can be guaranteed by the bearing localizability of the target formation \cite[Lemma~3]{zhao2015Maneuver}. Second, multiplying $K_i$ on both sides of \eqref{eq_bearingBasedControlLaw_singleIntegrator_maneuverPD} yields $\dot{\varepsilon}_i=-k_p\varepsilon_i$ where $\varepsilon_i= k_p\sum_{j\in\N_i} P_{g_{ij}^*}(p_i(t)-p_j(t))$ for $i\in\V_f$. It follows that $\varepsilon_i\rightarrow0$ as $t\rightarrow\infty$ for all $i\in\V_f$, and consequently $g_{ij}\rightarrow g_{ij}^*$ when the network is bearing localizable.

Under the action of the control laws \eqref{eq_bearingBasedControlLaw_singleIntegrator_maneuverPI} and \eqref{eq_bearingBasedControlLaw_singleIntegrator_maneuverPD}, the formation is able to perform translational and scaling formation maneuvers.
A translational maneuver means that all the agents move at a common velocity such that the formation translates as a rigid body.
A scaling maneuver means that the scale of the formation, which can be described by the distance from each agent to the formation centroid,
varies while the geometric pattern of the formation is preserved. In order to achieve the scaling maneuver, the leaders only needs to adjust the distances among them. One merit of the bearing-based control laws is that the desired maneuver is only known to the leaders and the followers are not required to access or estimate it.

\subsection*{\textbf{Bearing-Based Formation Control of Double Integrators}}

\begin{figure}
  \centering
  \subfloat[Generated formation maneuver trajectory (the dark area represents an obstacle).]{\includegraphics[width=\linewidth]{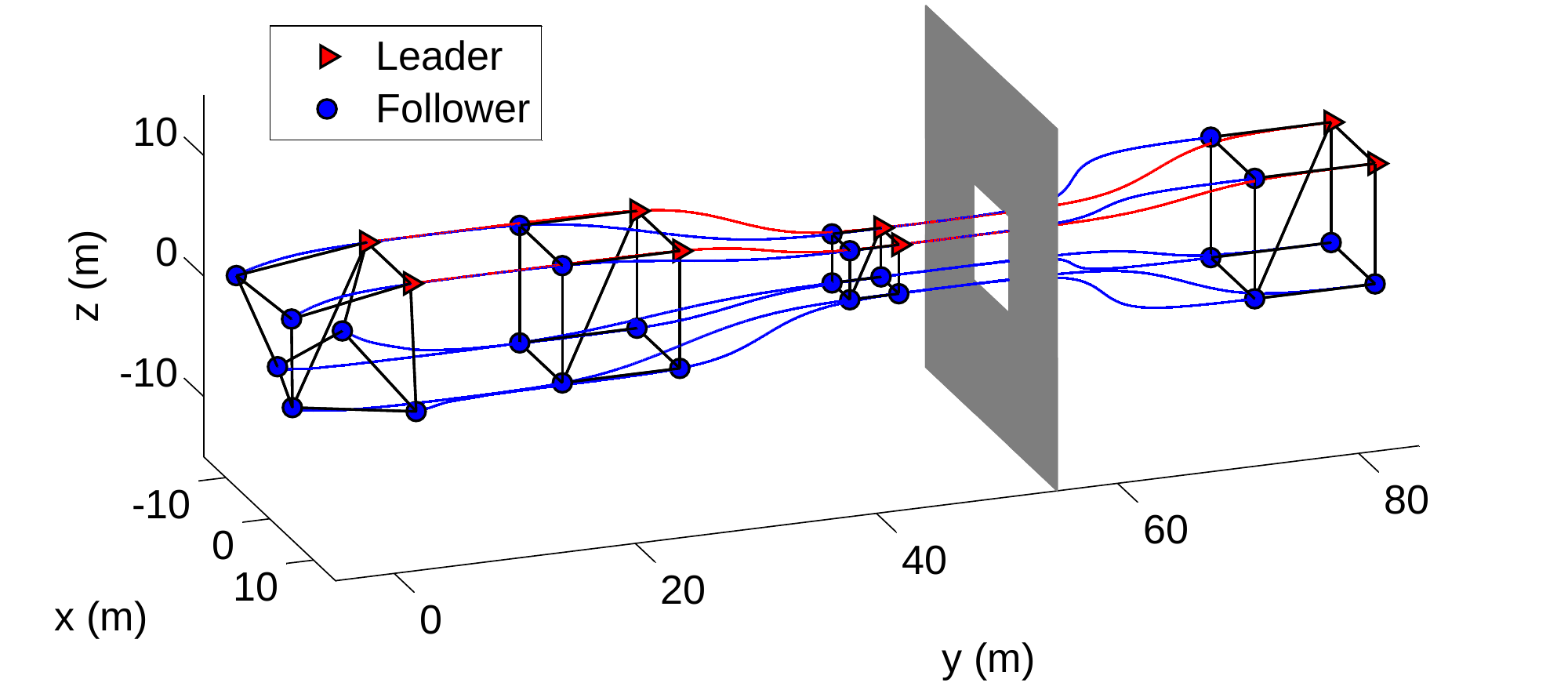}}\\
  \subfloat[Total bearing error of the trajectory, $\sum_{(i,j)\in\E}\|g_{ij}(t)-g_{ij}^*\|$.]{\includegraphics[width=\linewidth]{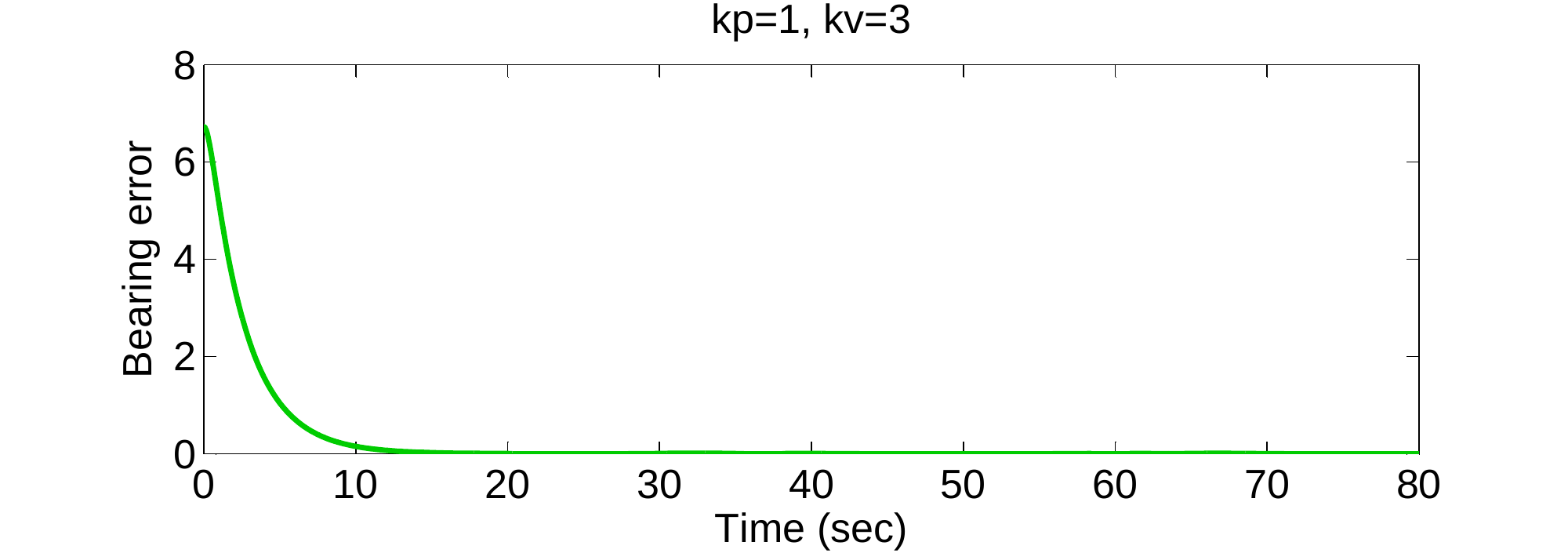}}\\
  \caption{Simulation example to demonstrate the bearing formation maneuvering control law in \eqref{eq_bearingBasedControlLaw_DoubleIntegrator_Maneuver_varyingVelocity}. The target formation in the example is a three-dimensional cube with two leaders and six followers.
    The translation and scale of the formation can continuously vary while the formation pattern is maintained as desired.
    This example demonstrates that formation scale control can be used for obstacle avoidance such as passing through narrow passages.}
  \label{fig_sim_3DTransVaryingVelocity}
\end{figure}

Consider the case where the dynamics of each mobile agent can be modeled as a double integrator
\begin{align*}
\dot{p}_i(t)&=v_i(t),\\
\dot{v}_i(t)&=u_i(t),
\end{align*}
where $u_i(t)$ is the acceleration input to be designed.
If the velocities of the leaders are constant, the bearing-based formation control problem can be solved by \cite{zhao2015Maneuver}
\begin{align}\label{eq_bearingBasedControlLaw_DoubleIntegrator_Maneuver_ConstantVelocity}
\dot{p}_i(t)&=v_i(t),\nonumber\\
\dot{v}_i(t)&=-\sum_{j\in\N_i}P_{g_{ij}^*}\Big[k_p(p_i(t)-p_j(t)) +k_v(v_i(t)-v_j(t))\Big],
\end{align}
where $i \in \V_f$ and $k_p, k_v$ are positive constant control gains.
Under control law \eqref{eq_bearingBasedControlLaw_DoubleIntegrator_Maneuver_ConstantVelocity}, the target formation is globally stable if it is bearing localizable.

If the velocities of the leaders are time-varying, the following control law requiring acceleration feedback can be used to track time-varying target formations \cite{zhao2015Maneuver},
\begin{align}\label{eq_bearingBasedControlLaw_DoubleIntegrator_Maneuver_varyingVelocity}
\dot{p}_i(t)&=v_i(t),\nonumber\\
\dot{v}_i(t)&=K_i^{-1}\sum_{j\in\N_i} P_{g^*_{ij}}\Big[-k_p(p_i(t)-p_j(t))\Big.\nonumber\\
&\qquad\qquad\Big.-k_v(v_i(t)-v_j(t))+\dot{v}_j(t)\Big],
\end{align}
where $i\in\V_f$ and $K_i=\sum_{j\in\N_i}P_{g_{ij}^*}$. The nonsingularity of $K_i$ for any $i\in\V_f$ is guaranteed by the bearing localizability of the target formation \cite[Lemma~3]{zhao2015Maneuver}.
Under control law \eqref{eq_bearingBasedControlLaw_DoubleIntegrator_Maneuver_varyingVelocity}, the target formation is globally stable if and only if it is bearing localizable.
A simulation example is given in Figure~\ref{fig_sim_3DTransVaryingVelocity} to demonstrate control law \eqref{eq_bearingBasedControlLaw_DoubleIntegrator_Maneuver_varyingVelocity}. In practice, absolute acceleration can be measured by each agent using, for example, inertial measurement units, and then transmitted to their neighbors by wireless communication. Due to measurement errors and transmission delays, the acceleration measurement is corrupted by errors. However, since the system is linear, bounded acceleration errors would cause bounded tracking errors.
Bearing-based formation control in the presence of some other problems including input disturbance, input saturation, and collision avoidance have been addressed in \cite{zhao2015Maneuver}.

\subsection*{\textbf{Bearing-Based Formation Control of Unicycles}}

\begin{figure}
  \centering
    \includegraphics[width=0.8\linewidth]{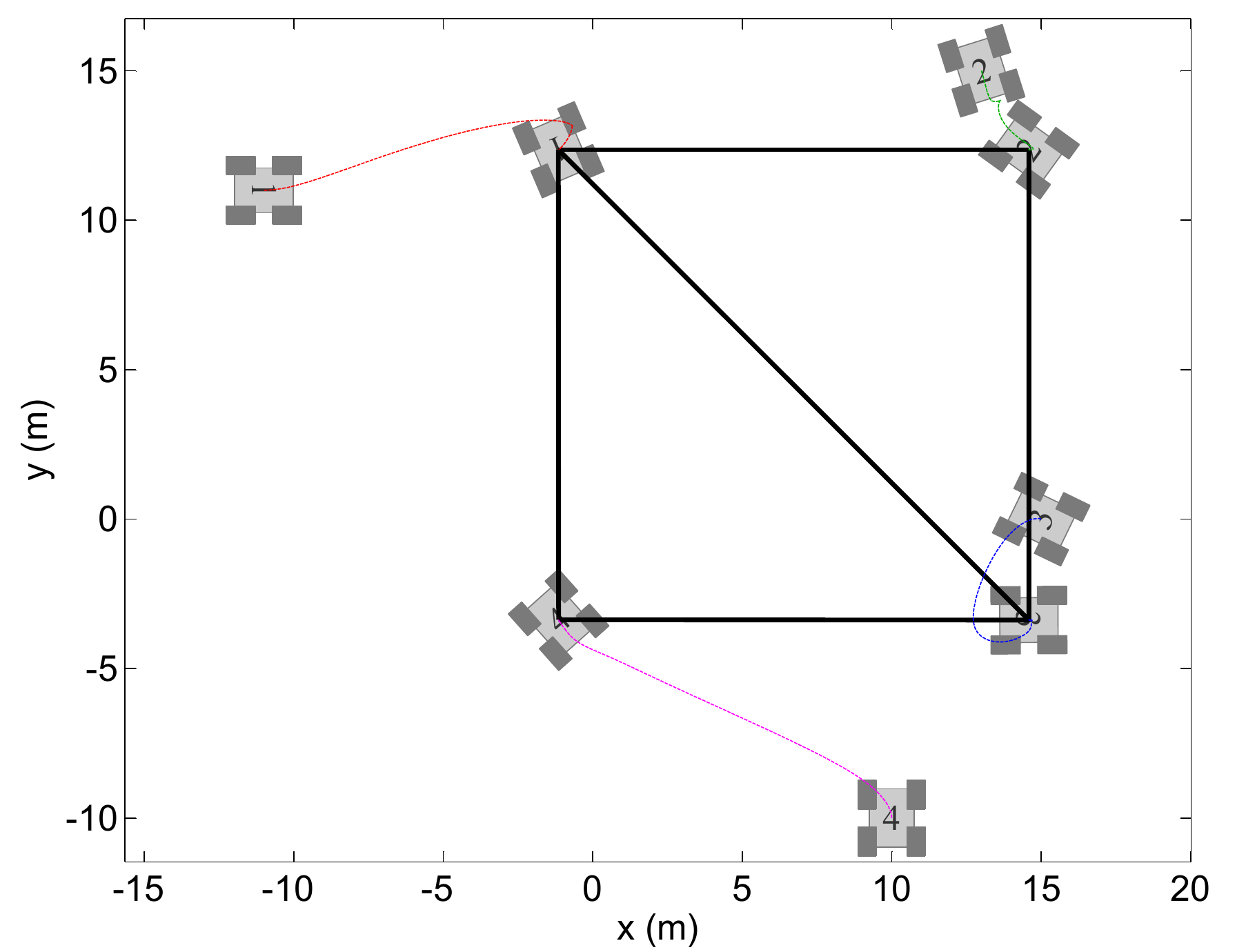}
  \caption{Simulation example to demonstrate the control law in \eqref{eq_bearingFormationUnicycle}. In this example, there are four unicycle agents whose initial positions and heading angles are chosen randomly. As can be seen, the formation converges to the target formation whose square geometric pattern is defined by five bearing vectors.}
  \label{fig_sim_bearingFormation_unicycle}
\end{figure}

Suppose the dynamics of agent $i\in\V$ can be described by the unicycle model
\begin{align*}
\dot{x}_i&=v_i\cos\theta_i, \nonumber\\
\dot{y}_i&=v_i\sin\theta_i, \nonumber\\
\dot{\theta}_i&=w_{i},
\end{align*}
where $p_i=[x_i,y_i]^T\in\R^2$ is the coordinate of agent $i$, $\theta_i\in\mathcal{S}^1$ is the heading angle, and $v_i\in\R$ and $w_{i}\in\R$ are the linear and angular velocities to be designed. Here, $\mathcal{S}^1$ is the one-dimensional manifold on the unit circle. The bearing-based formation control law for unicycles is \cite{zhao2017Constraints}
\begin{align}\label{eq_bearingFormationUnicycle}
v_i&=[\cos\theta_i \; \sin\theta_i]\sum_{j\in\N_i} P_{g_{ij}^*} (p_j(t)-p_i(t)), \nonumber\\
w_{i}&= [-\sin\theta_i\; \cos\theta_i]\sum_{j\in\N_i} P_{g_{ij}^*} (p_j(t)-p_i(t)).
\end{align}
When there are no leaders, control law \eqref{eq_bearingFormationUnicycle} ensures global stability in the sense that $g_{ij}(t)$ converges to either $g_{ij}^*$ or $-g_{ij}^*$ as $t\rightarrow\infty$ given any initial values of $p_i(0)$ and $\theta_i(0)$ if the target formation is infinitesimally bearing rigid \cite{zhao2017Constraints}. The final value of $\theta_i$ is not specified in the control law.
A simulation example is shown in Figure~\ref{fig_sim_bearingFormation_unicycle}.

\section*{\textbf{Bearing-Only Formation Control}}\label{section_bearingOnlyFormationControl}

This section introduces the theory of bearing-only formation control, which studies how to steer a group of agents to achieve a bearing-constrained target formation using bearing-only measurements. Suppose the target formation is specified by constant bearing constraints $\{g_{ij}^*\}_{(i,j)\in\E}$, and there are no leaders. The control objective is to control the positions of the agents $\{p_i(t)\}_{i \in \V}$ such that $g_{ij}(t)\rightarrow g_{ij}^*$ for all ${(i,j)\in\E}$ as $t\rightarrow\infty$.
All the bearings are expressed in a common reference frame.

\begin{figure}
  \centering
\includegraphics[width=\linewidth]{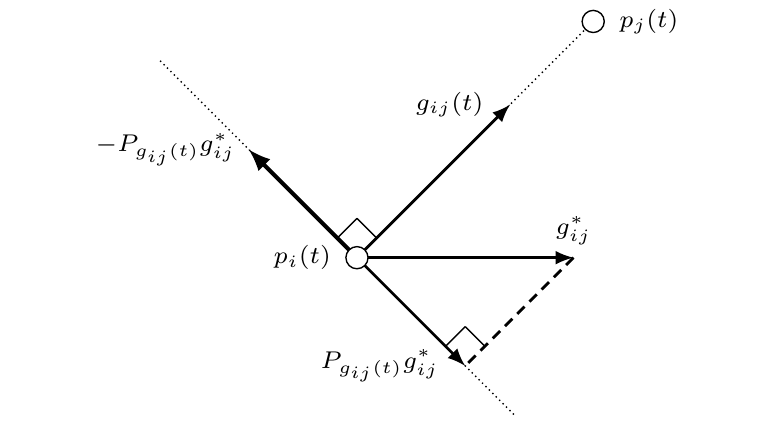}
  \caption{The geometric interpretation of the bearing-only control law in \eqref{eq_bearingOnlyControlLawElement}. Since the control term $-P_{g_{ij}}g_{ij}^*$ is perpendicular to the bearing $g_{ij}$, the control law aims to reduce the bearing error of $g_{ij}(t)$ while maintaining the distance between agents $i$ and $j$.}
  \label{fig_bearingOnlyControlLawGeometricMeaning}
\end{figure}

\begin{figure}
  \centering
\includegraphics[width=0.7\linewidth]{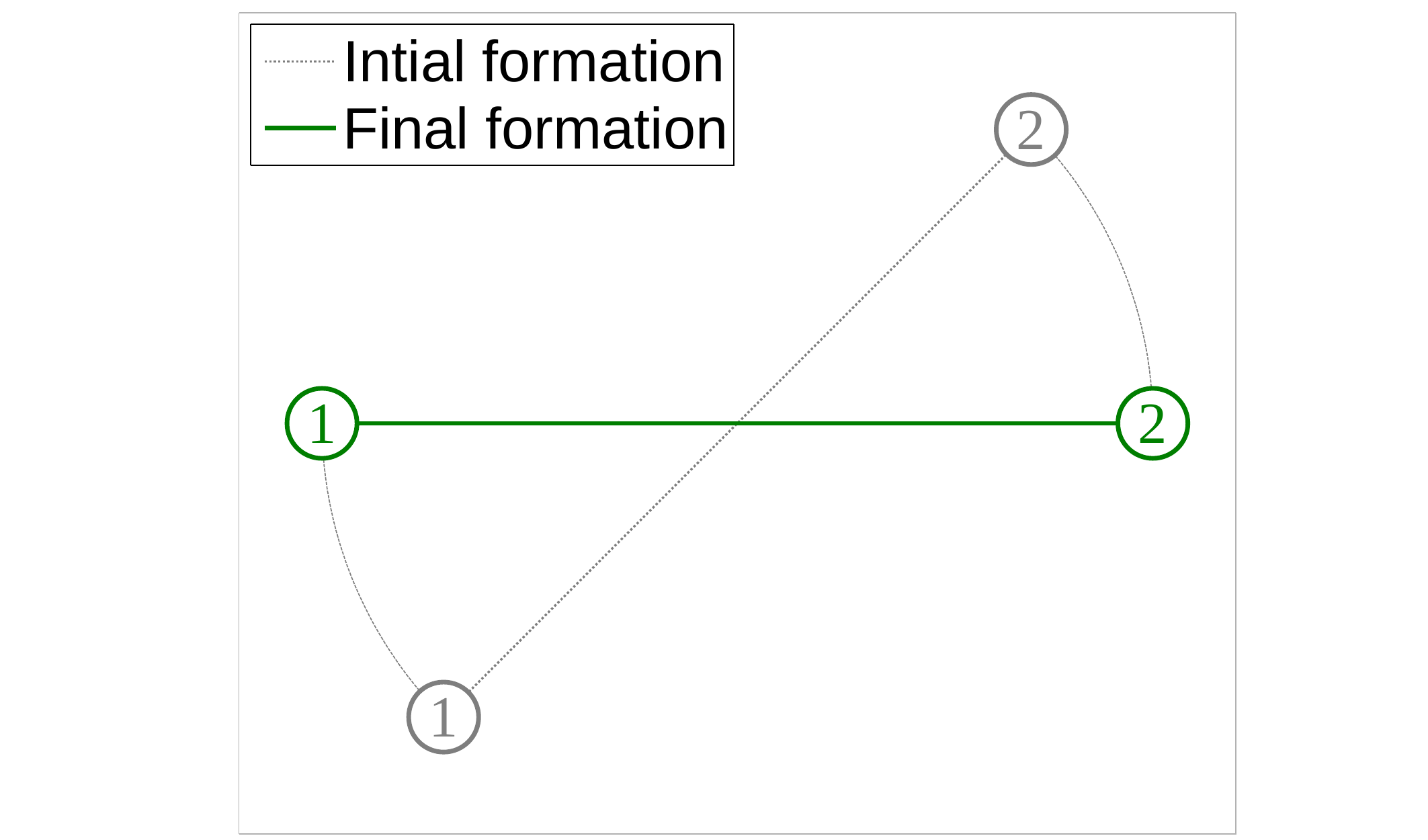}
  \caption{Simulation example to demonstrate the bearing-only formation control law in \eqref{eq_bearingOnlyControlLawElement}.
  In this example, the formation has two agents and one edge. In the target formation, the bearings are in the horizontal direction; that is $g_{12}^*=-g_{21}^*=[1,0]^T$.
  The initial formation (the dotted line in the figure) does not fulfil the desired bearings. Under the control law in \eqref{eq_bearingOnlyControlLawElement}, the formation converges to the desired one (the solid line in the figure). 
  Note that the velocity of each agent is always perpendicular to the bearing and hence the two agents move on a circle centered at their midpoint. As a result, the centroid and scale of the formation are invariant.}
  \label{fig_simExampleDemo}
\end{figure}

\begin{figure}
  \centering
  \subfloat[Initial configuration (grey circle) and final desired formation (blue circles).]{\includegraphics[width=.75\linewidth]{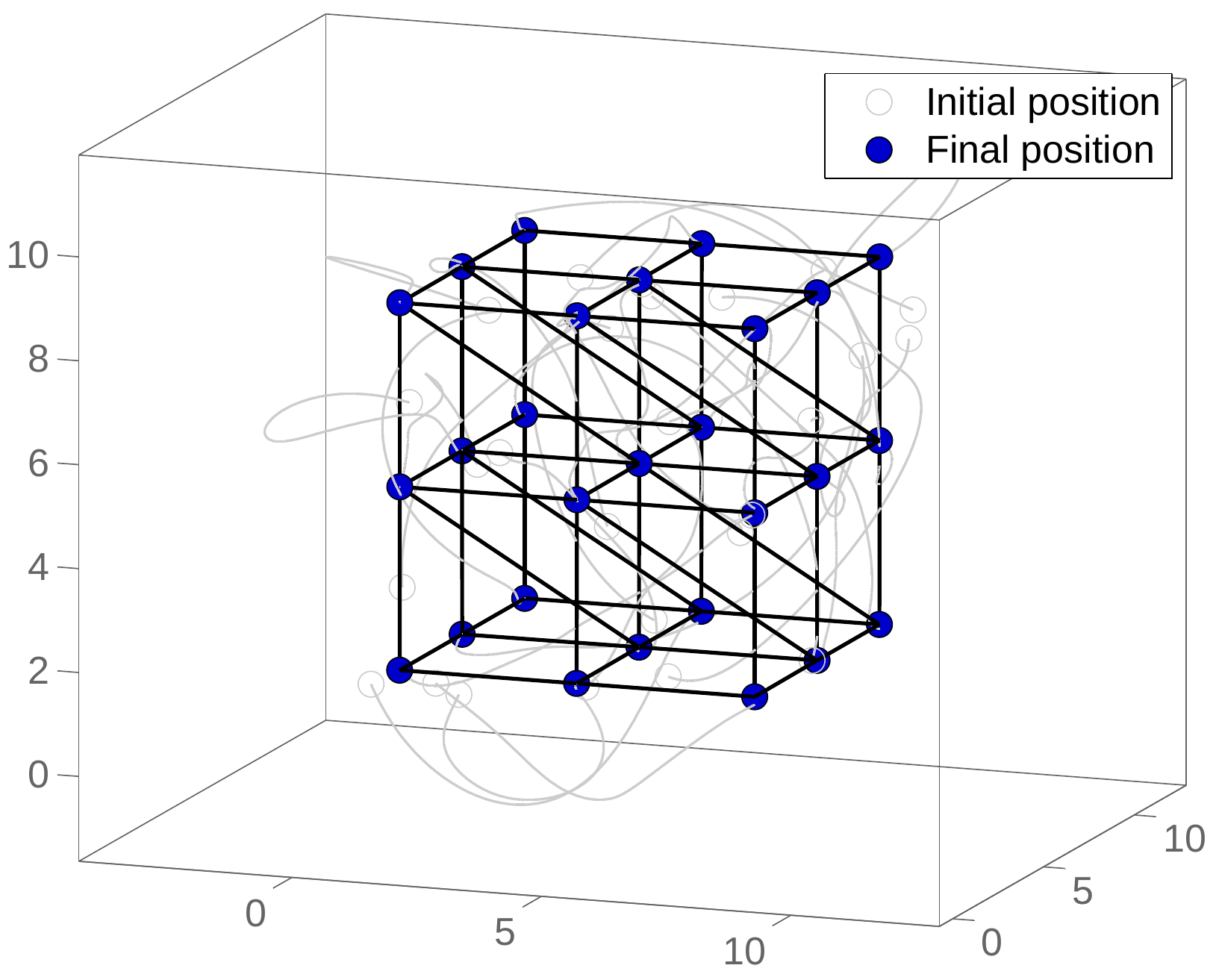}}\\
  \subfloat[Plot of the bearing error, $\sum_{(i,j)\in\E}\|g_{ij}(t)-g_{ij}^*\|$.]{\includegraphics[width=.75\linewidth]{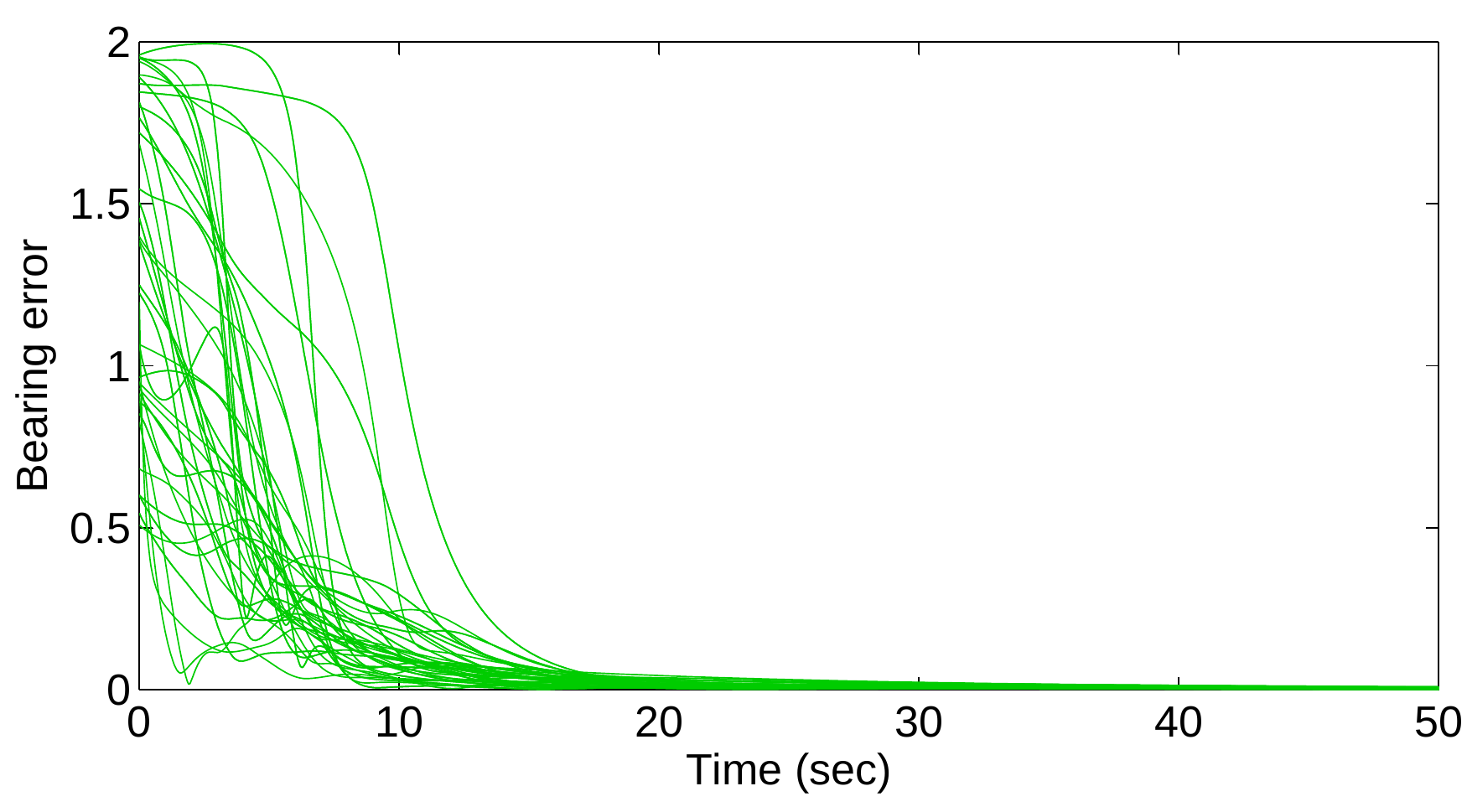}}
  \caption{Simulation example for the bearing control law in \eqref{eq_bearingOnlyControlLawElement} in three-dimensional space. In this example, the formation has 27 nodes and 62 edges. For the target formation, the rank of the bearing rigidity matrix, which equals the rank of the bearing Laplacian matrix, is $3n-4=77$. As a result, the target formation is infinitesimally bearing rigid and hence the control law \eqref{eq_bearingOnlyControlLawElement} is almost globally stable. As can be seen, given a random initial configuration, the target formation is achieved and the bearing errors converge to zero.}
  \label{fig_sim_bearingOnlyFormation}
\end{figure}

The following nonlinear control law, proposed in \cite{zhao2014TACBearing}, can be used to solve the bearing-only formation control problem,
\begin{align}\label{eq_bearingOnlyControlLawElement}
    \dot{p}_i(t) = - \sum_{j\in\N_i} P_{g_{ij}(t)}g_{ij}^*, \quad i\in\V,
\end{align}
where $P_{g_{ij}(t)}=I_d-g_{ij}(t)g_{ij}^T (t)$.
The geometric interpretation of the control law is illustrated in Figure~\ref{fig_bearingOnlyControlLawGeometricMeaning}.
Some properties of the control law are highlighted below.
First, the control of each agent only requires bearing measurements and does not require distance or position estimation.
Second, the control input of \eqref{eq_bearingOnlyControlLawElement} is always bounded as $\|\dot{p}_i(t)\|\le \sum_{j\in\N_i} \|P_{g_{ij}(t)}\|\|g_{ij}^*\|=|\N_i|$, since $\|P_{g_{ij}(t)}\|=\|g_{ij}^*\|=1$.
Third, the centroid and scale of the formation are invariant under the control law \cite[Theorem~9]{zhao2014TACBearing}.
Here, the centroid is defined as the average position of the agents and the scale is defined as the standard deviation of the distances from the agents to the centroid.
Simulation examples are given in Figures~\ref{fig_simExampleDemo} and \ref{fig_sim_bearingOnlyFormation} to demonstrate control law \eqref{eq_bearingOnlyControlLawElement}.

System \eqref{eq_bearingOnlyControlLawElement} is nonlinear and almost globally stable if the target formation is infinitesimally bearing rigid \cite[Theorem~11]{zhao2014TACBearing}.
The term ``almost'' is due to the fact that there are two isolated equilibriums of the error dynamics, one is desired and the other is undesired.
At the desired equilibrium, the bearings are equal to the desired values; that is $g_{ij}=g_{ij}^*$ for $(i,j)\in\E$.
At the undesired equilibrium, the bearings are opposite to the desired values; that is $g_{ij}=-g_{ij}^*$ for $(i,j)\in\E$.
The formations at the two equilibriums have the same centroid and scale but opposite bearings.
The almost global stability means that the formation would converge to the desired equilibrium unless the initial formation lies exactly on the undesired equilibrium, which can be shown to be an unstable equilibrium.

Control law \eqref{eq_bearingOnlyControlLawElement} is a modified gradient-descent control law. In particular, consider the following objective function,
\begin{align*}
\phi_1=\frac{1}{2}\sum_{(i,j)\in\E} \|g_{ij}-g_{ij}^*\|^2=\sum_{(i,j)\in\E} (1-g_{ij}^T g_{ij}^*).
\end{align*}
The objective function is equal to zero if and only if $g_{ij}=g_{ij}^*$ for all $(i,j)\in\E$. The corresponding gradient-descent control law is
\begin{align}\label{eq_bearingOnlyGradientControlLaw}
\dot{p}_i(t) = - \sum_{j\in\N_i} \frac{1}{\|e_{ij}(t)\|} P_{g_{ij}(t)} g_{ij}^*, \quad i\in\V.
\end{align}
The two-dimensional version of control law \eqref{eq_bearingOnlyGradientControlLaw} was first proposed in \cite{bishopconf2011rigid}. This control law requires both bearing and distance measurements. Removing the distance term $\|e_{ij}(t)\|$ in \eqref{eq_bearingOnlyGradientControlLaw} yields the bearing-only formation control law in \eqref{eq_bearingOnlyControlLawElement}.

An optimization-based approach for bearing-only formation control can be found in \cite{Tron2016CDC,Tron2016CSM}, where a bearing-only control law is proposed as
\begin{align}\label{eq_bearingOnlyControl_Tron}
\dot{p}_i(t) = \sum_{j\in\N_i} (g_{ij}(t)-g_{ij}^*), \quad i\in\V.
\end{align}
This is a gradient-descent control law with the corresponding objective function as
\begin{align*}
\phi_2=\frac{1}{4}\sum_{(i,j)\in\E} \|e_{ij}\|\|g_{ij}-g_{ij}^*\|^2=\frac{1}{2}\sum_{(i,j)\in\E} \|e_{ij}\|(1-g_{ij}^Tg_{ij}^*).
\end{align*}
Since $\phi_2$ contains $\|e_{ij}\|$, $\phi_2$ is zero when $g_{ij}=g_{ij}^*$ or $e_{ij}=0$. As a result, the scale of the formation always decreases under the action of control law \eqref{eq_bearingOnlyControl_Tron}. Simulation shows that this control law may steer all the agents to the same position given certain initial conditions. To avoid this problem, leaders must be introduced \cite{Tron2016CDC}.

\section*{\textbf{Conclusions and Future Directions}}\label{section_discussion}

This article presented a review of the bearing rigidity theory and its applications in distributed formation control and network localization for multi-agent systems.
Motivated by the fact that many existing approaches rely on measurement assumptions that may be difficult to realize under certain circumstances, this article demonstrated how to utilize bearing-only sensors, such as cameras or sensor arrays, to solve the problems of formation control and network localization. The article discussed three specific problems including bearing-based network localization, bearing-based formation control, and bearing-only formation control.

As a newly emerged research area, bearing-based control and estimation is far from being fully explored. Many important problems in this area remain unsolved. One key assumption for the results presented in this article is that the underlying graph is undirected, which means any pair of neighbors must be able to access each other's information. Since this assumption may not be valid in some practical tasks, it is important to study the case of directed graphs. When the graph is directed, the control and estimation problem would become more complicated because undesired equilibriums may emerge, as observed in \cite{zhao2015CDC}. Similar problems also exist in distance-based formation control \cite{Hendrickx2007IJRNC,Yu2009SIAM,Summers2011TAC}. Despite the resent progress on bearing-only formation control for some special directed graphs \cite{Mukherjee2017Cyclic,MinhIJRNC2017}, the problem for general directed graphs remains an important challenge in this area.

\begin{figure}
  \centering
  \subfloat[Initial formation.]{\includegraphics[width=0.5\linewidth]{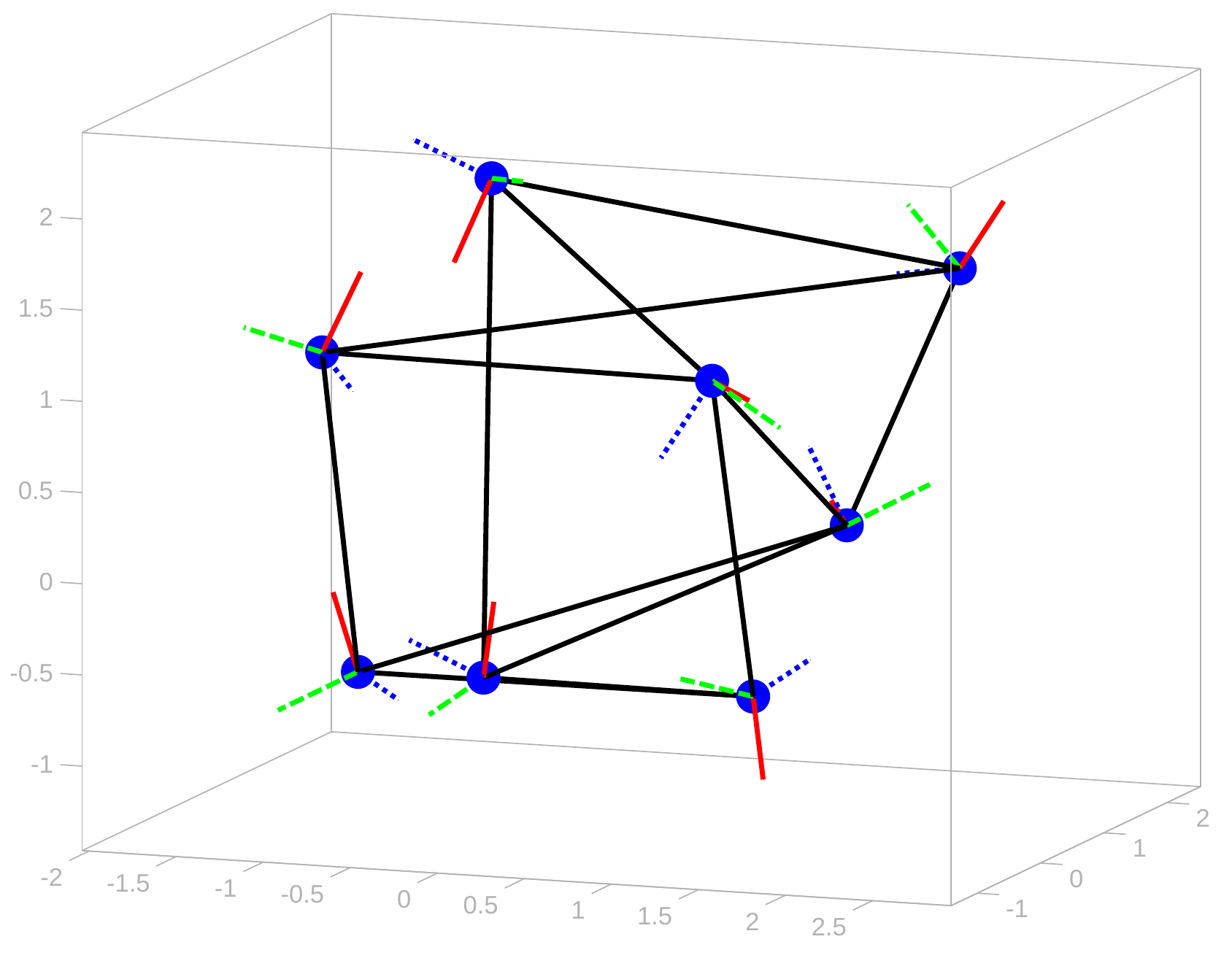}}
  \subfloat[Final formation.]{\includegraphics[width=0.5\linewidth]{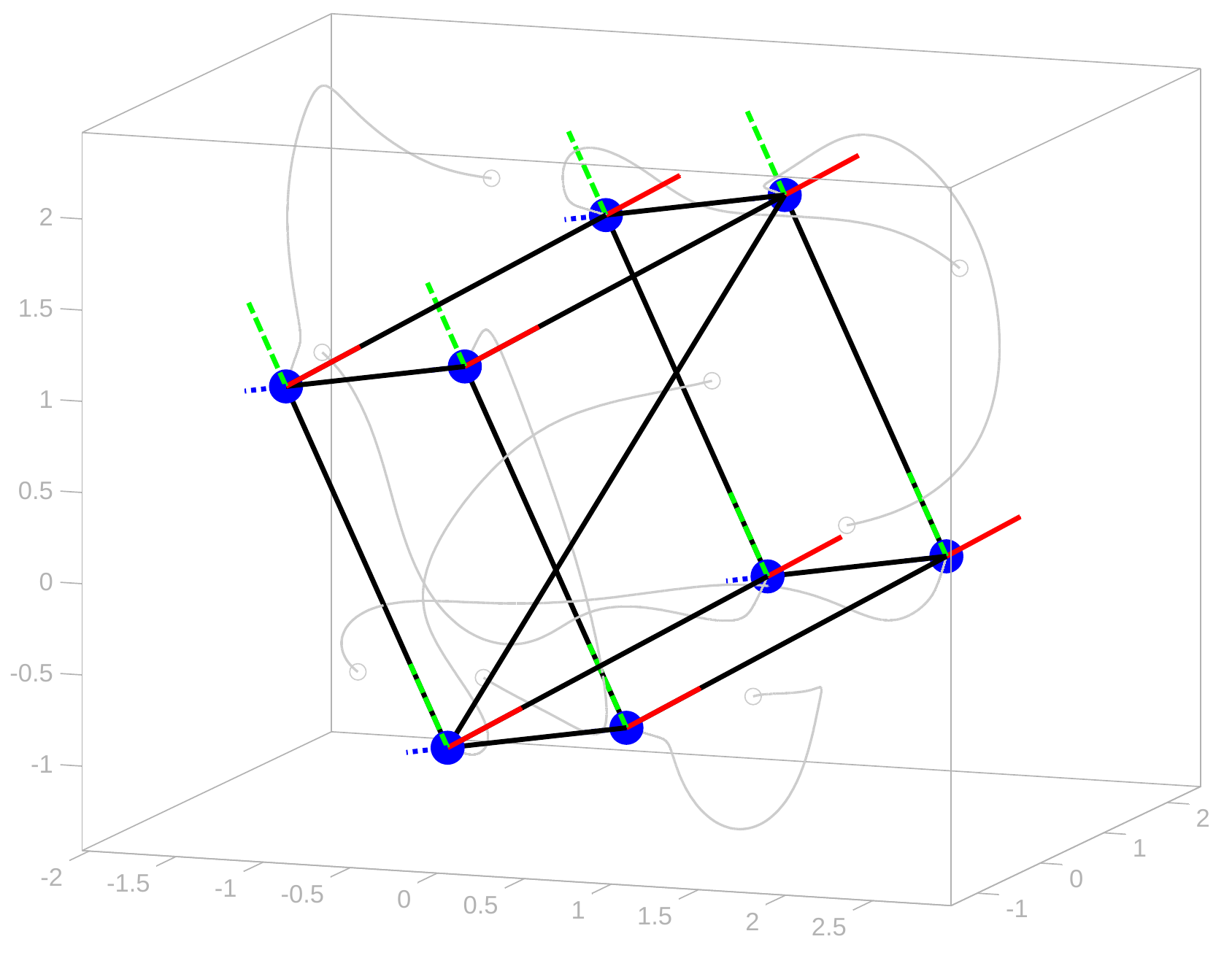}}
  \caption{Simulation example for bearing-only formation control without a global reference frame. The control law is given in \cite[Equation~(19)]{zhao2014TACBearing}. In this example, the formation has 8 nodes and 13 edges. The target formation is a three-dimensional cube that is infinitesimally bearing rigid. The control is based on inter-neighbor bearings expressed in each agent's local reference frames. The orientations of the agents are synchronized in the final formation.}
  \label{fig_sim_bearingOnlyFormationNoGlobal}
\end{figure}

Another key assumption for the results addressed in this article is that all bearings must be measured in a global reference frame. Global reference frames, however, may not be accessible to each agent in some environments such as indoors. It is important to study how to achieve control or estimation when bearings are measured in each agent's local reference frames. One potential approach is to estimate or synchronize the orientations of the local reference frames \cite{Tron2016CSM,zhao2014TACBearing}.  This approach has been applied to adapt the bearing-only formation control law in \eqref{eq_bearingOnlyControlLawElement} to use locally measured bearings \cite[Section~IV]{zhao2014TACBearing}, and a simulation example is given in Figure~\ref{fig_sim_bearingOnlyFormationNoGlobal}. This is also a general approach for many types of formation control and network localization tasks in the absence of global reference frames \cite{KKOh2014TAC,Ahn2016Orientation}. However, distributed orientation estimation or synchronization requires each agent to obtain their neighbors' relative orientations, which are usually difficult to measure in practice.
Other potential approaches that do not require an orientation estimation may be based on bearing rigidity in the special Euclidean group $SE(n)$ \cite{zelazo2014SE2Rigidity,zelazo2015CDC,Schiano2016,Michieletto2016CDC,Schiano2017} or complex Laplacian \cite{Lin2016TSP,Han2016Scale}. A brief introduction to bearing rigidity in $SE(2)$ is given in ``\nameref{sidebar_SE2Rigidity}.'' Nevertheless, the formation control strategies provided for $SE(2)$ frameworks still require additional sensing \cite{zelazo2015CDC}, and a complete theory for bearing-only formation control is still unsolved.

In addition to network localization and formation control, many other tasks may also be achieved with bearing-only measurements such as bearing-only rendezvous \cite{YuTACBearingOnlyRendezvous,ZhengRonghao2013SCL,Grzymisch2015JGDC,Kriegleder2015ICRA,zhao2017CCC}, and bearing-only target tracking \cite{Loizou2007CDC,Deghat2014TAC,Dimos2014US,ZhengRonghao2015Automatica,Trinh2016Guidance,Mengbin2017TAES} though the analysis of these tasks may not rely on the bearing rigidity theory.

The bearing rigidity theory and its application for formation control and network localization is strongly motivated by the sensing mediums available to distributed and multi-agent systems.  This work contributed to a broader theory of cooperative control and estimation for networked systems and hopes to serve as a starting point for both practitioners and theoreticians in this community.

\newpage
\clearpage
\bibliography{myOwnPub,zsyReferenceAll} 

\begin{thebibliography}{10}

\bibitem{Cao2013OverviewMultiagent}
Y.~Cao, W.~Yu, W.~Ren, and G.~Chen, ``An overview of recent progress in the
  study of distributed multi-agent coordination,'' {\em IEEE Transactions on
  Industrial Informatics}, vol.~9, no.~1, pp.~427--438, 2013.

\bibitem{Oh2015Automatica}
K.-K. Oh, M.-C. Park, and H.-S. Ahn, ``A survey of multi-agent formation
  control,'' {\em Automatica}, vol.~53, pp.~424--440, March 2015.

\bibitem{Akyildiz2002}
I.~F. Akyildiz, W.~Su, Y.~Sankarasubramaniam, and E.~Cayirci, ``A survey on
  sensor networks,'' {\em IEEE Communications Magazine}, vol.~40, pp.~102--114,
  Aug 2002.

\bibitem{AspnesTMC2006}
J.~Aspnes, T.~Eren, D.~K. Goldenberg, A.~S. Morse, W.~Whiteley, Y.~R. Yang,
  B.~D.~O. Anderson, and P.~N. Belhumeur, ``A theory of network localization,''
  {\em IEEE Transactions on Mobile Computing}, vol.~12, no.~5, pp.~1663--1678,
  2006.

\bibitem{Mao2007NetworkLocalization}
G.~Mao, B.~Fidan, and B.~D.~O. Anderson, ``Wireless sensor network localization
  techniques,'' {\em Comupter Networks}, vol.~51, pp.~2529--2553, 2007.

\bibitem{Barooah2007}
P.~Barooah and J.~P. Hespanha, ``Estimation on graphs from relative
  measurements,'' {\em IEEE Control Systems Magzine}, no.~1, pp.~57--74, 2007.

\bibitem{MayiBook}
Y.~Ma, S.~Soatto, J.~Kosecka, and S.~Sastry, {\em An Invitation to 3D Vision}.
\newblock New York: Springer, 2004.

\bibitem{Hartley1997Triangulation}
R.~I. Hartley and P.~Sturm, ``Triangulation,'' {\em Computer vision and image
  understanding}, vol.~68, no.~2, pp.~146--157, 1997.

\bibitem{nima2009TR}
N.~Moshtagh, N.~Michael, A.~Jadbabaie, and K.~Daniilidis, ``Vision-based,
  distributed control laws for motion coordination of nonholonomic robots,''
  {\em IEEE Transactions on Robotics}, vol.~25, pp.~851--860, August 2009.

\bibitem{Tron2016CSM}
R.~Tron, J.~Thomas, G.~Loianno, K.~Daniilidis, and V.~Kumar, ``A distributed
  optimization framework for localization and formation control: applications
  to vision-based measurements,'' {\em IEEE Control Systems Magazine}, vol.~36,
  no.~4, pp.~22--44, 2016.

\bibitem{BearingTracking1999}
A.~Farina, ``Target tracking with bearings-only measurements,'' {\em Signal
  Processing}, vol.~78, pp.~61--68, 1999.

\bibitem{Liuwei2017}
Y.-Y. Dong, C.-X. Dong, W.~Liu, H.~Chen, and G.-Q. Zhao, ``2-{D} {DOA}
  estimation for {L}-shaped array with array aperture and snapshots extension
  techniques,'' {\em IEEE Signal Processing letters}, vol.~24, no.~4,
  pp.~495--499, 2017.

\bibitem{Stacey2016TAC}
G.~Stacey and R.~Mahony, ``A passivity-based approach to formation control
  using partial measurements of relative position,'' {\em IEEE Transactions on
  Automatic Control}, vol.~61, no.~2, pp.~538--543, 2016.

\bibitem{BearingObservability1993}
K.~Becker, ``Simple linear theory approach to {TMA} observability,'' {\em IEEE
  Transactions on Aerospace and Electronic Systems}, vol.~29, no.~2,
  pp.~575--578, 1993.

\bibitem{Anderson2014TAC}
M.~Deghat, I.~Shames, B.~D.~O. Anderson, and C.~Yu, ``Localization and
  circumnavigation of a slowly moving target using bearing measurements,'' {\em
  IEEE Transactions on Automatic Control}, vol.~59, no.~8, pp.~2182--2188,
  2014.

\bibitem{Dimos2014US}
J.~O. Swartling, I.~Shames, K.~H. Johansson, and D.~V. Dimarogonas,
  ``Collective circumnavigation,'' {\em Unmanned Systems}, vol.~2, no.~3,
  pp.~219--229, 2014.

\bibitem{ZhengRonghao2015Automatica}
R.~Zheng, Y.~Liu, and D.~Sun, ``Enclosing a target by nonholonomic mobile
  robots with bearing-only measurements,'' {\em Automatica}, vol.~53,
  pp.~400--407, 2015.

\bibitem{Mengbin2017TAES}
M.~Ye, B.~D.~O. Anderson, and C.~Yu, ``Bearing-only measurement
  self-localization, velocity consensus and formation control,'' {\em IEEE
  Transactions on Aerospace and Electronic Systems}, vol.~52, no.~2,
  pp.~575--586, 2017.

\bibitem{OlfatiTAC2004}
R.~Olfati-Saber and R.~M. Murray, ``Consensus problems in networks of agents
  with switching topology and time-delays,'' {\em IEEE Transactions on
  Automatic Control}, vol.~49, no.~9, pp.~1520--1533, 2004.

\bibitem{LinZhiyun2005TAC}
Z.~Lin, B.~Francis, and M.~Maggiore, ``Necessary and sufficient graphical
  conditions for formation control,'' {\em IEEE Transactions on Automatic
  Control}, vol.~50, pp.~121--127, January 2005.

\bibitem{Ren2007CSM}
W.~Ren, R.~W. Beard, and E.~M. Atkins, ``Information consensus in multivehicle
  cooperative control,'' {\em IEEE Control Systems Magazine}, vol.~27,
  pp.~71--82, April 2007.

\bibitem{LiZhongkui2010}
Z.~Li, Z.~Duan, G.~Chen, and L.~Huang, ``Consensus of multiagent systems and
  synchronization of complex networks: A unified viewpoint,'' {\em IEEE
  Transactions on Circuits and Systems I: Regular Papers}, vol.~57, no.~1,
  pp.~213--224, 2010.

\bibitem{Whiteley1999ParallelDraw}
B.~Servatius and W.~Whiteley, ``Constraining plane configurations in
  computer-aided design: Combinatorics of directions and lengths,'' {\em SIAM
  Journal on Discrete Mathematics}, vol.~12, no.~1, pp.~136--153, 1999.

\bibitem{bishopconf2011rigid}
A.~N. Bishop, ``Stabilization of rigid formations with direction-only
  constraints,'' in {\em Proceedings of the 50th IEEE Conference on Decision
  and Control and European Control Conference}, (Orlando, FL, USA),
  pp.~746--752, December 2011.

\bibitem{Eren2012IJC}
T.~Eren, ``Formation shape control based on bearing rigidity,'' {\em
  International Journal of Control}, vol.~85, no.~9, pp.~1361--1379, 2012.

\bibitem{zelazo2014SE2Rigidity}
D.~Zelazo, A.~Franchi, and P.~R. Giordano, ``Rigidity theory in {SE(2)} for
  unscaled relative position estimation using only bearing measurements,'' in
  {\em Proceedings of the 2014 European Control Conference}, (Strasbourgh,
  France), pp.~2703--2708, June 2014.

\bibitem{TronRigidComponent}
R.~Tron, L.~Carlone, F.~Dellaert, and K.~Daniilidis, ``Rigid components
  identification and rigidity control in bearing-only localization using the
  graph cycle basis,'' in {\em Proceedings of the 2015 American Control
  Conference}, (Chicago, IL, USA), pp.~3911--3918, 2015.

\bibitem{zhao2014TACBearing}
S.~Zhao and D.~Zelazo, ``Bearing rigidity and almost global bearing-only
  formation stabilization,'' {\em IEEE Transactions on Automatic Control},
  vol.~61, no.~5, pp.~1255--1268, 2016.

\bibitem{Asimow1978Rigidty}
L.~Asimow and B.~Roth, ``The rigidity of graphs,'' {\em Transactions of the
  American Mathematical Society}, vol.~245, pp.~279--289, November 1978.

\bibitem{ConnellyBook}
R.~Connelly and S.~D. Guest, {\em Frameworks, Tensegrities and Symmetry:
  Understanding Stable Structures}.
\newblock 2015.
\newblock Available online at
  \url{http://www.math.cornell.edu/~web7510/framework.pdf} (last accessed
  November 2017).

\bibitem{Hendrickson1992SIAM}
B.~Hendrickson, ``Conditions for unique graph realizations,'' {\em SIAM Journal
  on Computing}, vol.~21, no.~1, pp.~65--84, 1992.

\bibitem{Connelly2005Generic}
R.~Connelly, ``Generic global rigidity,'' {\em Discrete \& Computational
  Geometry}, vol.~33, pp.~549--563, 2005.

\bibitem{Jacobs1997}
D.~Jacobs, ``An algorithm for two-dimensional rigidity percolation: The pebble
  game,'' {\em Journal of Computational Physics}, vol.~137, pp.~346--365, Nov.
  1997.

\bibitem{WhiteleyHenneberg1985}
T.-S. Tay and W.~Whiteley, ``Generating isostatic frameworks,'' {\em Structural
  Topology}, vol.~11, pp.~21--69, 1985.

\bibitem{Whiteley2005Pseudotriangulation}
R.~Haas, D.~Orde, G.~Rote, F.~Santos, B.~Servatius, H.~Servatius, D.~Souvaine,
  I.~Streinu, and W.~Whiteley, ``Planar minimally rigid graphs and
  pseudo-triangulations,'' {\em Computational Geometry}, vol.~31, no.~1--2,
  pp.~31--61, 2005.

\bibitem{Jackson2007NotesRigidity}
B.~Jackson, ``Notes on the rigidity of graphs,'' tech. rep., Queen Mary
  University of London, 2007.

\bibitem{Anderson2008CSM}
B.~D.~O. Anderson, C.~Yu, B.~Fidan, and J.~Hendrickx, ``Rigid graph control
  architectures for autonomous formations,'' {\em IEEE Control Systems
  Magazine}, vol.~28, pp.~48--63, December 2008.

\bibitem{Krick2009IJC}
L.~Krick, M.~E. Broucke, and B.~A. Francis, ``Stabilization of infinitesimally
  rigid formations of multi-robot networks,'' {\em International Journal of
  Control}, vol.~82, no.~3, pp.~423--439, 2009.

\bibitem{oh2013IJRNC}
K.-K. Oh and H.-S. Ahn, ``Distance-based undirected formations of
  single-integrator and double-integrator modeled agents in $n$-dimensional
  space,'' {\em International Journal of Robust and Nonlinear Control},
  vol.~24, pp.~1809--1820, August 2014.

\bibitem{Tian2013Global}
Y.-P. Tian and Q.~Wang, ``Global stabilization of rigid formations in the
  plane,'' {\em Automatica}, vol.~49, pp.~1436--1441, May 2013.

\bibitem{zelazo2015Rigidity}
D.~Zelazo, A.~Franchi, and P.~R. Giordano, ``Distributed rigidity maintenance
  control with range-only measurements for multi-robot systems,'' {\em The
  International Journal of Robotics Research}, vol.~34, no.~1, pp.~105--128,
  2015.

\bibitem{Mou2016TAC}
S.~Mou, M.-A. Belabbas, A.~S. Morse, Z.~Sun, and B.~D.~O. Anderson,
  ``Undirected rigid formations are problematic,'' {\em IEEE Transactions on
  Automatic Control}, vol.~61, no.~10, pp.~2821--2836, 2016.

\bibitem{ChenXudong2016SIAM}
X.~Chen, M.-A. Belabbas, and T.~Ba\c{s}ar, ``Global stabilization of
  triangulated formations,'' {\em SIAM Journal on Optimization and Control},
  vol.~55, no.~1, pp.~172--199, 2017.

\bibitem{ZhiyongAutomatica2016}
Z.~Sun, M.-C. Park, B.~D.~O. Anderson, and H.-S. Ahn, ``Distributed
  stabilization control of rigid formations with prescribed orientation,'' {\em
  Automatica}, vol.~78, pp.~250--257, 2017.

\bibitem{Marina2016TR}
H.~G. de~Marina, B.~Jayawardhana, and M.~Cao, ``Distributed rotational and
  translational maneuvering of rigid formations and their applications,'' {\em
  IEEE Transactions on Robotics}, vol.~32, no.~3, pp.~684--697, 2016.

\bibitem{Eren2004}
T.~Eren, O.~K. Goldenberg, W.~Whiteley, Y.~R. Yang, A.~S. Morse, B.~D.~O.
  Anderson, and P.~Belhumeur, ``Rigidity, computation, and randomization in
  network localization,'' in {\em INFOCOM 2004. Twenty-third Annual Joint
  Conference of the IEEE Computer and Communications Societies}, (Pasadena,
  USA), pp.~2673--2684, 2004.

\bibitem{eren2003}
T.~Eren, W.~Whiteley, A.~S. Morse, P.~N. Belhumeur, and B.~D.~O. Anderson,
  ``Sensor and network topologies of formations with direction, bearing and
  angle information between agents,'' in {\em Proceedings of the 42nd IEEE
  Conference on Decision and Control}, (Hawaii, USA), pp.~3064--3069, December
  2003.

\bibitem{BishopTAES2009}
A.~N. Bishop, B.~D.~O. Anderson, B.~Fidan, P.~N. Pathirana, and G.~Mao,
  ``Bearing-only localization using geometrically constrained optimization,''
  {\em IEEE Transactions on Aerospace and Electronic Systems}, vol.~45, no.~1,
  pp.~308--320, 2009.

\bibitem{Piovan2013Automatica}
G.~Piovan, I.~Shames, B.~Fidan, F.~Bullo, and B.~D.~O. Anderson, ``On frame and
  orientation localization for relative sensing networks,'' {\em Automatica},
  vol.~49, pp.~206--213, January 2013.

\bibitem{Shames2013TAC}
I.~Shames, A.~N. Bishop, and B.~D.~O. Anderson, ``Analysis of noisy
  bearing-only network localization,'' {\em IEEE Transactions on Automatic
  Control}, vol.~58, pp.~247--252, January 2013.

\bibitem{ZhuGuangwei2014Automatica}
G.~Zhu and J.~Hu, ``A distributed continuous-time algorithm for network
  localization using angle-of-arrival information,'' {\em Automatica}, vol.~50,
  pp.~53--63, January 2014.

\bibitem{Lin2016TSP}
Z.~Lin, T.~Han, R.~Zheng, and M.~Fu, ``Distributed localization for 2-{D}
  sensor networks with bearing-only measurements under switching topologies,''
  {\em IEEE Transactions on Signal Processing}, vol.~64, no.~23,
  pp.~6345--6359, 2016.

\bibitem{Bishop2015Relaxed}
A.~N. Bishop, M.~Deghat, B.~D.~O. Anderson, and Y.~Hong, ``Distributed
  formation control with relaxed motion requirements,'' {\em International
  Journal of Robust and Nonlinear Control}, vol.~25, pp.~3210--3230, 2015.

\bibitem{zhao2015ECC}
S.~Zhao and D.~Zelazo, ``Bearing-based distributed control and estimation in
  multi-agent systems,'' in {\em Proceedings of the 2015 European Control
  Conference}, (Linz, Austria), pp.~2207--2212, July 2015.

\bibitem{zhao2015Maneuver}
S.~Zhao and D.~Zelazo, ``Translational and scaling formation maneuver control
  via a bearing-based approach,'' {\em IEEE Transactions on Control of Network
  Systems}, vol.~4, no.~3, pp.~429--438, 2017.

\bibitem{Fathian2016ACC}
K.~Fathian, D.~I. Rachinskii, M.~W. Spong, and N.~R. Gans, ``Globally
  asymptotically stable distributed control for distance and bearing based
  multi-agent formations,'' in {\em Proceedings of the 2016 American Control
  Conference}, (Boston, MA, USA), pp.~4642--4648, 2016.

\bibitem{Han2016Scale}
Z.~Han, L.~Wang, Z.~Lin, and R.~Zheng, ``Formation control with size scaling
  via a complex {Laplacian}-based approach,'' {\em IEEE Transactions on
  Cybernetics}, vol.~46, no.~10, pp.~2348--2359, 2016.

\bibitem{Coogan2012Scale}
S.~Coogan and M.~Arcak, ``Scaling the size of a formation using relative
  position feedback,'' {\em Automatica}, vol.~48, pp.~2677--2685, October 2012.

\bibitem{bishop2010SCL}
M.~Basiri, A.~N. Bishop, and P.~Jensfelt, ``Distributed control of triangular
  formations with angle-only constraints,'' {\em Systems \& Control Letters},
  vol.~59, pp.~147--154, 2010.

\bibitem{Franchi2012IJRR}
A.~Franchi, C.~Masone, V.~Grabe, M.~Ryll, H.~H. B{u}lthoff, and P.~R. Giordano,
  ``Modeling and control of {UAV} bearing formations with bilateral high-level
  steering,'' {\em The International Journal of Robotics Research}, vol.~31,
  no.~12, pp.~1504--1525, 2012.

\bibitem{Cornejo2013IJRR}
A.~Cornejo, A.~J. Lynch, E.~Fudge, S.~Bilstein, M.~Khabbazian, and J.~McLurkin,
  ``Scale-free coordinates for multi-robot systems with bearing-only sensors,''
  {\em The International Journal of Robotics Research}, vol.~32, no.~12,
  pp.~1459--1474, 2013.

\bibitem{zhao2013SCLDistribued}
S.~Zhao, F.~Lin, K.~Peng, B.~M. Chen, and T.~H. Lee, ``Distributed control of
  angle-constrained cyclic formations using bearing-only measurements,'' {\em
  Systems \& Control Letters}, vol.~63, no.~1, pp.~12--24, 2014.

\bibitem{zhao2013IJCFinite}
S.~Zhao, F.~Lin, K.~Peng, B.~M. Chen, and T.~H. Lee, ``Finite-time
  stabilization of cyclic formations using bearing-only measurements,'' {\em
  International Journal of Control}, vol.~87, no.~4, pp.~715--727, 2014.

\bibitem{Eric2014ACC}
E.~Schoof, A.~Chapman, and M.~Mesbahi, ``Bearing-compass formation control: a
  human-swarm interaction perspective,'' in {\em Proceedings of the 2014
  American Control Conference}, (Portland, USA), pp.~3881--3886, June 2014.

\bibitem{Tron2016CDC}
R.~Tron, J.~Thomas, G.~Loianno, K.~Daniilidis, and V.~Kumar, ``Bearing-only
  formation control with auxiliary distance measurements, leaders, and
  collision avoidance,'' in {\em Proceedings of the 55th Conference on Decision
  and Control}, (Las Vegas, USA), pp.~1806--1813, 2016.

\bibitem{zhao2015NetLocalization}
S.~Zhao and D.~Zelazo, ``Localizability and distributed protocols for
  bearing-based network localization in arbitrary dimensions,'' {\em
  Automatica}, vol.~69, pp.~334--341, 2016.

\bibitem{BookGodsilGraph}
C.~Godsil and G.~Royle, {\em Algebraic Graph Theory}.
\newblock New York: Springer, 2001.

\bibitem{zhao2015CDC}
S.~Zhao and D.~Zelazo, ``Bearing-based formation stabilization with directed
  interaction topologies,'' in {\em Proceedings of the 54th IEEE Conference on
  Decision and Control}, (Osaka, Japan), pp.~6115--6120, December 2015.

\bibitem{zhao2017CDCLaman}
S.~Zhao, Z.~Sun, D.~Zelazo, M.~H. Trinh, and H.-S. Ahn, ``Laman graphs are
  generically bearing rigid in arbitrary dimensions,'' in {\em Proceedings of
  the 56th IEEE Conference on Decision and Control}, (Melbourne, Australia),
  December 2017.
\newblock in press (available online at
  \url{https://arxiv.org/abs/1703.04035}).

\bibitem{Laman1970}
G.~Laman, ``On graphs and rigidity of plane skeletal structures,'' {\em Journal
  of Engineering Mathematics}, vol.~4, no.~4, pp.~331--340, 1970.

\bibitem{zhao2015MSC}
S.~Zhao and D.~Zelazo, ``Bearing-based formation maneuvering,'' in {\em
  Proceedings of the 2015 IEEE Multi-Conference on Systems and Control},
  (Sydney, Australia), pp.~658--663, September 2015.

\bibitem{zhao2017Constraints}
S.~Zhao, D.~V. Dimarogonas, Z.~Sun, and D.~Bauso, ``A general approach to
  coordination control of mobile agents with motion constraints,'' {\em IEEE
  Transactions on Automatic Control}.
\newblock in press (DOI: 10.1109/TAC.2017.2750924).

\bibitem{Hendrickx2007IJRNC}
J.~Hendrickx, B.~D.~O. Anderson, J.~Delvenne, and V.~Blondel, ``Directed graphs
  for the analysis of rigidity and persistence in autonomous agent systems,''
  {\em International Journal of Robust and Nonlinear Control}, vol.~17,
  no.~10-11, pp.~960--981, 2007.

\bibitem{Yu2009SIAM}
C.~Yu, B.~D.~O. Anderson, A.~S. Dasgupta, and B.~Fidan, ``Control of minimally
  persistent formations in the plane,'' {\em SIAM Journal on Control and
  Optimization}, vol.~48, pp.~206--233, Februray 2009.

\bibitem{Summers2011TAC}
T.~H. Summers, C.~Yu, S.~Dasgupta, and B.~D.~O. Anderson, ``Control of
  minimally persistent leader-remote-follower and coleader formations in the
  plane,'' {\em IEEE Transactions on Automatic Control}, vol.~56, no.~12,
  pp.~2778--2792, 2011.

\bibitem{Mukherjee2017Cyclic}
D.~Mukherjee, M.-H. Trinh, D.~Zelazo, and H.-S. Ahn, ``Bearing-only cyclic
  pursuit in 2-{D} for capture of moving target,'' in {\em 57th Israel Annual
  Conference on Aerospace Sciences}, (Haifa, Israel), 2017.

\bibitem{MinhIJRNC2017}
M.-H. Trinh, D.~Mukherjee, D.~Zelazo, and H.-S. Ahn, ``Formations on directed
  cycles with bearing-only measurements,'' {\em International Journal of Robust
  and Nonlinear Control}, pp.~1--23, 2017.
\newblock in press (DOI: 10.1002/rnc.3921).

\bibitem{KKOh2014TAC}
K.-K. Oh and H.-S. Ahn, ``Formation control and network localization via
  orientation alignment,'' {\em IEEE Transactions on Automatic Control},
  vol.~59, pp.~540--545, February 2014.

\bibitem{Ahn2016Orientation}
B.-H. Lee and H.-S. Ahn, ``Distributed formation control via global orientation
  estimation,'' {\em Automatica}, vol.~73, pp.~125--129, 2016.

\bibitem{zelazo2015CDC}
D.~Zelazo, A.~Franchi, and P.~R. Giordano, ``Formation control using a {SE}(2)
  rigidity theory,'' in {\em Proceedings of the 54th IEEE Conference on
  Decision and Control}, (Osaka, Japan), pp.~6121--6126, 2015.

\bibitem{Schiano2016}
F.~Schiano, A.~Franchi, D.~Zelazo, and P.~R. Giordano, ``{A Rigidity-Based
  Decentralized Bearing Formation Controller for Groups of Quadrotor UAVs},''
  in {\em Proceedings of IEEE/RSJ International Conference on Intelligent
  Robots and Systems}, (Daejeon, Korea), pp.~5099--5106, 2016.

\bibitem{Michieletto2016CDC}
G.~Michieletto, A.~Cenedese, and A.~Franchi, ``Bearing rigidity theory in
  {SE}(3),'' in {\em Proceedings of the 55th IEEE Conference on Decision and
  Control}, (5950--5955), pp.~83--92, 2016.

\bibitem{Schiano2017}
F.~Schiano and P.~G. Robuffo, ``Bearing rigidity maintenance for formations of
  quadrotor {UAVs},'' in {\em Proceedings of IEEE International Conference on
  Robotics and Automation}, (Singapore), 2017.

\bibitem{YuTACBearingOnlyRendezvous}
J.~Yu, S.~M. LaValle, and D.~Liberzon, ``Rendezvous without coordinates,'' {\em
  IEEE Transactions on Automatic Control}, vol.~57, no.~2, pp.~421--434, 2012.

\bibitem{ZhengRonghao2013SCL}
R.~Zheng and D.~Sun, ``Rendezvous of unicycles: A bearings-only and perimeter
  shortening approach,'' {\em Systems \& Control Letters}, vol.~62,
  pp.~401--407, May 2013.

\bibitem{Grzymisch2015JGDC}
J.~Grzymisch and W.~Fichter, ``Optimal rendezvous guidance with enhanced
  bearings-only observability,'' {\em Journal of Guidance, Control, and
  Dynamics}, vol.~38, no.~6, pp.~1131--1139, 2015.

\bibitem{Kriegleder2015ICRA}
M.~Kriegleder, S.~T. Digumarti, R.~Oung, and R.~D'Andrea, ``Rendezvous with
  bearing-only information and limited sensing range,'' in {\em Proceedings of
  the 2015 IEEE International Conference on Robotics and Automation}, (Seattle,
  Washington), pp.~5941--5947, 2015.

\bibitem{zhao2017CCC}
S.~Zhao and R.~Zheng, ``Flexible bearing-only rendezvous control of mobile
  robots,'' in {\em Proceedings of the 36th Chinese Control Conference},
  (Dalian, China), pp.~8051--8056, July 2017.

\bibitem{Loizou2007CDC}
S.~G. Loizou and V.~Kumar, ``Biologically inspired bearing-only navigation and
  tracking,'' in {\em Proceedings of the 46th IEEE Conference on Decision and
  Control}, (New Orleans, LA, USA), pp.~1386--1391, 2007.

\bibitem{Deghat2014TAC}
M.~Deghat, I.~Shames, A.~N. Bishop, B.~D.~O. Anderson, and C.~Yu,
  ``Localization and circumnavigation of a slowly moving target using bearing
  measurements,'' {\em IEEE Transactions on Automatic Control}, vol.~59,
  pp.~2182--2188, August 2014.

\bibitem{Trinh2016Guidance}
M.-H. Trinh, G.-H. Ko, V.-H. Pham, K.-K. Oh, and H.-S. Ahn, ``Guidance using
  bearing-only measurements with three beacons in the plane,'' {\em Control
  Engineering Practice}, vol.~51, pp.~81--91, 2016.

\end{thebibliography}
\bibliographystyle{ieeetr}

\setcounter{section}{0}
\newpage
\renewcommand{\thetheorem}{S\arabic{theorem}}
\setcounter{theorem}{0}
\renewcommand{\thelemma}{S\arabic{lemma}}
\setcounter{lemma}{0}
\renewcommand{\theproposition}{S\arabic{proposition}}
\setcounter{proposition}{0}
\renewcommand{\thedefinition}{S\arabic{definition}}
\setcounter{definition}{0}
\renewcommand{\thefigure}{S\arabic{figure}}
\setcounter{figure}{0}
\renewcommand{\theequation}{S\arabic{equation}}
\setcounter{equation}{0}
\renewcommand{\thetable}{S\arabic{table}}
\setcounter{table}{0}

%

\newpage
\section*{Notations for Networks and Formations}
\label{sidebar_notationsforNetworksFormations}

Given a network of $n$ nodes in $\R^d$ where $n\ge2,d\ge2$, let the position of node $i$ be $p_i\in\R^d$ and the configuration of the points be $p=[p_1^T,\dots,p_n^T]^T\in\mathbb{R}^{dn}$.
The interaction among the nodes is described by a graph $\mathcal{G}=(\mathcal{V},\mathcal{E})$ which consists of a vertex set $\mathcal{V}=\{1,\dots,n\}$ and an edge set $\mathcal{E}\subseteq \mathcal{V} \times \mathcal{V}$. If $(i,j)\in\E$, node $i$ receives information from node $j$, and node $j$ is called adjacent to $i$.
The set of neighbors of vertex $i$ is denoted as $\mathcal{N}_i=\{j \in \mathcal{V}: (i,j)\in \mathcal{E}\}$.
This article focuses on undirected graphs where $(i,j)\in\E\Leftrightarrow (j,i)\in\E$.
Let $m$ be the number of undirected edges in the graph.
An orientation of an undirected graph is the assignment of a direction to each edge.
An oriented graph is an undirected graph together with an orientation.
The incidence matrix $H\in\mathbb{R}^{m\times n}$ of an oriented graph is the $\{0,\pm1\}$-matrix with rows indexed by edges and columns by vertices.

A network, denoted as $(\G,p)$, is $\G$ with its vertex $i\in\V$ mapped to $p_i$. Network may be called as formation in the context of formation control.
For a network $(\G,p)$, define the edge and bearing vectors for $(i,j)\in\E$ as $e_{ij}= p_j-p_i$ and $g_{ij}= e_{ij}/\|e_{ij}\|$, respectively.
Here $g_{ij}$ is the unit vector pointing from $p_i$ to $p_j$. It represents the relative bearing of $p_i$ with respect to $p_j$.
Note that $e_{ij}=-e_{ji}$ and $g_{ij}=-g_{ji}$.
Consider an orientation of the graph $\G$ and suppose $(i,j)$ corresponds to the $k$th edge in the oriented graph. Then the edge and bearing vectors may be reexpressed as $e_{k}= p_j-p_i$ and $g_{k}= {e_{k}}/{\|e_{k}\|}$ where $k\in\{1,\dots,m\}$.
Denote $e=[e_1^T ,\dots,e_m^T ]^T$ and $g=[g_1^T ,\dots,g_m^T ]^T$.
Note that $e=(H\otimes I_d)p$ where $\otimes$ denotes the Kronecker product.
In this article, $\Null(\cdot)$ and $\Range(\cdot)$ denote the null and range spaces of a matrix, respectively.
Denote $\one_n\triangleq[1,\dots,1]^T\in\R^n$. Let $\|\cdot\|$ be the Euclidian norm of a vector or the spectral norm of a matrix, and $I_d\in\R^{d\times d}$ be the identity matrix.

\clearpage
\newpage

\renewcommand{\thefigure}{S\arabic{figure}}
\setcounter{figure}{0}
\section*{Key Definitions in Bearing Rigidity Theory}
\label{sidebar_bearingRigidity}

\begin{definition}[Bearing Equivalency]
    Two networks $(\G,p)$ and $(\G,p')$ are \emph{bearing equivalent} if $P_{(p_i-p_j)}(p_i'-p_j')=0$ for all $(i,j)\in\E$.
\end{definition}

\begin{definition}[Bearing Congruency]
    Two networks $(\G,p)$ and $(\G,p')$ are \emph{bearing congruent} if $P_{(p_i-p_j)}(p_i'-p_j')=0$ for all $i,j\in\V$.
\end{definition}

\begin{definition}[Bearing Rigidity]
    A network $(\G,p)$ is \emph{bearing rigid} if there exists a constant $\epsilon>0$ such that any network $(\G,p')$ that is bearing equivalent to $(\G,p)$ and satisfies $\|p'-p\|<\epsilon$ is also bearing congruent to $(\G,p)$.
\end{definition}

\begin{definition}[Global Bearing Rigidity]\label{definition_globalParallelRigidity}
    A network $(\G,p)$ is \emph{globally bearing rigid} if an arbitrary network that is bearing equivalent to $(\G,p)$ is also bearing congruent to $(\G,p)$.
\end{definition}

Consider an oriented graph where the inter-neighbor bearings can be expressed by $\{g_k\}_{k=1}^m$.
Define the \emph{bearing function} $F_B: \R^{dn}\rightarrow\R^{dm}$ as
\begin{align*}
    F_B(p)= [g_1^T , \dots, g_m^T ]^T \in\R^{dm}.
\end{align*}
The \emph{bearing rigidity matrix} is defined as the Jacobian of the bearing function
\begin{align}\label{eq_rigidityMatrixDefinition}
    R_B(p) = \frac{\partial F_B(p)}{\partial p}\in\R^{dm\times dn}.
\end{align}
A matrix-vector form $R_B(p)$ is
\begin{align*}
R_B(p)=\mydiag(P_{g_{1}}/\|e_1\|,\dots,P_{g_m}/\|e_m\|)(H\otimes I_d).
\end{align*}
Let $\delta p\in\R^{dn}$ be a variation of the configuration $p$.
If $R_B(p)\delta p=0$, then $\delta{p}$ is called an \emph{infinitesimal bearing motion} of $(\G,p)$.
An infinitesimal bearing motion is called \emph{trivial} if it only corresponds to a translation and a scaling of the entire network.

\begin{definition}[Infinitesimal Bearing Rigidity]\label{definition_infinitesimalParallelRigid}
    A network is \emph{infinitesimally bearing rigid} if all the infinitesimal bearing motions are trivial.
\end{definition}

\begin{figure}[h]
  \centering
\includegraphics[width=\linewidth]{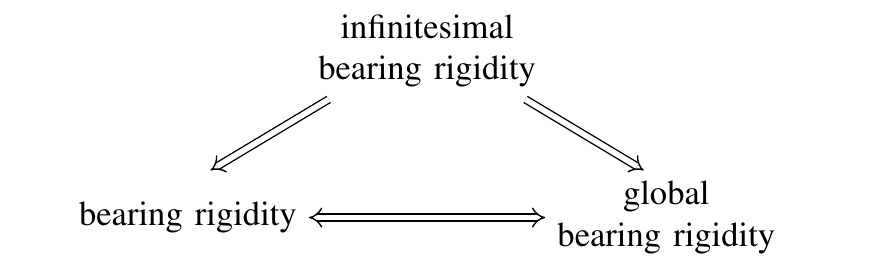}
  \caption{The relation between bearing rigidity, global bearing rigidity, and infinitesimal bearing rigidity. Infinitesimal bearing rigidity implies both bearing rigidity and global bearing rigidity. Global bearing rigidity and bearing rigidity imply each other.}
  \label{fig_relationBetweenTheRigiditiesBearing}
\end{figure}

The relation between bearing rigidity, global bearing rigidity, and infinitesimal bearing rigidity is illustrated in Figure~\ref{fig_relationBetweenTheRigiditiesBearing}.
Details of these notions can be found in \cite{zhao2014TACBearing}.

\newpage


\renewcommand{\thedefinition}{S\arabic{definition}}
\setcounter{definition}{0}
\section*{Key Definitions in Distance Rigidity Theory}
\label{sidebar_distanceRigidity}

\begin{definition}[Distance Equivalency]
    Two networks $(\G,p)$ and $(\G,p')$ are \emph{distance equivalent} if $\|p_i-p_j\|=\|p_i'-p_j'\|$ for all $(i,j)\in\E$.
\end{definition}

\begin{definition}[Distance Congruency]
    Two networks $(\G,p)$ and $(\G,p')$ are \emph{distance congruent} if $\|p_i-p_j\|=\|p_i'-p_j'\|$ for all $i,j\in\V$.
\end{definition}

\begin{definition}[Distance Rigidity]
    A network $(\G,p)$ is \emph{distance rigid} if there exists a constant $\epsilon>0$ such that any network $(\G,p')$ that is distance equivalent to $(\G,p)$ and satisfies $\|p'-p\|<\epsilon$ is also distance congruent to $(\G,p)$.
\end{definition}

\begin{definition}[Global Distance Rigidity]\label{definition_globalParallelRigidity}
    A network $(\G,p)$ is \emph{globally distance rigid} if an arbitrary network that is distance equivalent to $(\G,p)$ is also distance congruent to it.
\end{definition}

Consider an oriented graph where the inter-neighbor distances can be expressed by $\{\|e_k\|\}_{k=1}^m$.
Define the \emph{distance function} $F_D: \R^{dn}\rightarrow\R^{dm}$ as
\begin{align*}
    F_D(p)= [\|e_1\|^2, \dots, \|e_m\|^2]^T /2\in\R^{m}.
\end{align*}
The \emph{distance rigidity matrix} is defined as the Jacobian of the distance function
\begin{align}\label{eq_rigidityMatrixDefinition}
    R_D(p) = \frac{\partial F_D(p)}{\partial p}\in\R^{m\times dn}.
\end{align}
A matrix-vector form $R_D(p)$ is
\begin{align*}
R_D(p)=\mydiag(e_1^T,\dots,e_m^T )(H\otimes I_d).
\end{align*}
Let $\delta p\in\R^{dn}$ be a variation of the configuration $p$.
If $R_D(p)\delta p=0$, then $\delta{p}$ is called an \emph{infinitesimal distance motion} of $(\G,p)$.
An infinitesimal distance motion is called \emph{trivial} if it only corresponds to a translation and a rotation of the entire network.

\begin{definition}[Infinitesimal Distance Rigidity]\label{definition_infinitesimalParallelRigid}
    A network is \emph{infinitesimally distance rigid} if all the infinitesimal distance motions are trivial.
\end{definition}

\begin{figure}[h]
  \centering
  \includegraphics[width=\linewidth]{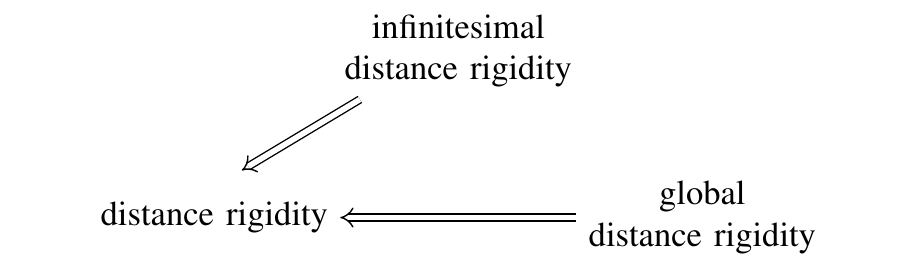}
  \caption{The relation between distance rigidity, global distance rigidity, and infinitesimal distance rigidity. Both infinitesimal and global distance rigidity imply distance rigidity. Infinitesimal and global distance rigidity do not imply each other.}
  \label{fig_relationBetweenTheRigiditiesDistance}
\end{figure}

The relation between distance rigidity, global distance rigidity, and infinitesimal distance rigidity is illustrated in Figure~\ref{fig_relationBetweenTheRigiditiesDistance}.
Details of the notions can be found in \cite{Asimow1978Rigidty,Hendrickson1992SIAM,Connelly2005Generic,Jackson2007NotesRigidity,ConnellyBook}.

\newpage
\section*{An Orthogonal Projection Matrix}\label{sidebar_projectionOperator}

\begin{figure}[h]
  \centering
\includegraphics[width=\linewidth]{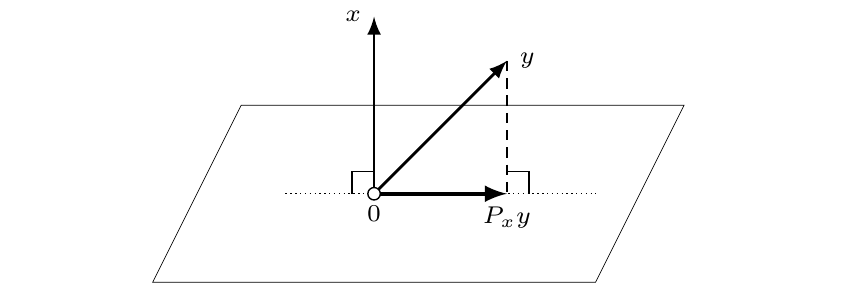}
  \caption{Illustration of the orthogonal projection matrix. Given any nonzero $x,y\in\R^d$, the vector $P_xy$ is the orthogonal projection of $y$ onto the orthogonal compliment of $x$.}
  \label{fig_PxGeometricMeaning}
\end{figure}

For any nonzero vector $x\in\R^d$ ($d\ge2$), define an orthogonal projection matrix as
\begin{align*}
    P(x) = I_d - \frac{x}{\|x\|}\frac{x^T }{\|x\|}\in\R^{d\times d}.
\end{align*}
For notational simplicity, denote $P_x = P(x)$.
The matrix $P_x$ is an orthogonal projection matrix that geometrically projects any vector onto the orthogonal compliment of $x$ (see Figure~\ref{fig_PxGeometricMeaning}).

Matrix $P_x$ satisfies $P_x^T =P_x$, $P_x^2=P_x$, and $\Null(P_x)=\myspan\{x\}$. This matrix is positive semi-definite with one eigenvalue equal to zero and $d-1$ eigenvalues equal to one. Other properties of $P_x$ are summarized as below.
\begin{enumerate}[(a)]
\item Any two nonzero vectors $x, y\in\R^d$ are parallel if and only if $P_x y=0$ \cite[Lemma~1]{zhao2014TACBearing}.
\item Any two unit vectors $x, y\in\R^d$ satisfy $x^T P_{y}x=y^T P_{x}y$ \cite[Lemma~8]{zhao2014TACBearing}.
\item For any nonzero vectors $x_1,\dots,x_m\in\R^d$ where $m\ge2,d\ge2$, the matrix $\sum_{i=1}^m P_{x_i}\in\R^{d\times d}$ is nonsingular if and only if at least two of $x_1,\dots,x_m$ are not collinear \cite[Lemma~3]{zhao2017CDCLaman}.
\item For any nonzero vector $x\in\R^2$, denote $x^\perp\in\R^2$ as a nonzero normal vector that satisfies $x^T  x^\perp=0$. Then $P_x=x^\perp(x^\perp)^T/\|x^\perp\|^2$. The proof follows from the fact that the matrix $A=[x/\|x\|,x^\perp/\|x^\perp\|]\in\R^{2\times 2}$ satisfies $A^TA=AA^T=I_2$.
\item For any two nonzero vectors $x, y\in\R^d$, if $\theta\in[0,\pi]$ is the angle between them so that $x^T  y=\|x\|\|y\|\cos\theta$, then $\|P_x-P_y\|=\sin\theta$ \cite[Lemma~5]{zhao2015NetLocalization}. This property has been used to analyze the perturbation of the orthogonal projection matrix.
\item If $x\in\R^3$ is a unit vector, then $P_x=-\sk{x}^2$, where
\begin{align*}
    \sk{x}=\left[
      \begin{array}{ccc}
        0 & -x_3 & x_2 \\
        x_3 & 0 & -x_1 \\
        -x_2 & x_1 & 0 \\
      \end{array}
    \right]\in\R^{3\times 3}
\end{align*}
is the skew-symmetric matrix associated with $x$ \cite[Theorem~2.11]{MayiBook}. This property has been used in \cite[Equation~(6)]{zhao2017Constraints}
\end{enumerate}
The orthogonal projection matrix plays an important role in the bearing rigidity theory and its applications.

\newpage
\section*{Bearing Laplacian of Networks}\label{sidebar_bearingLaplacian}

\begin{figure}[h]
  \centering
\includegraphics[width=\linewidth]{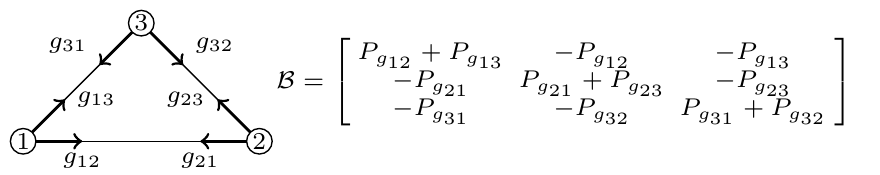}
  \caption{Example to demonstrate bearing Laplacian.  The network is the complete graph on three nodes.  The bearing Laplacian has the same structure as a weighted graph Laplacian matrix \cite{BookGodsilGraph} with the weights on each edge corresponding to the projection matrices $P_{g_{ij}}$.}
  \label{fig_demoBearingLaplacian}
\end{figure}

Given network $(G,p)$ with no collocated nodes, define the \emph{bearing Laplacian} $\L\in\R^{dn\times dn}$ as \cite{zhao2015NetLocalization}
\begin{align*}
[\L]_{ij} = \left\{
  \begin{array}{ll}
      {\bf 0}_{d \times d}, &i \neq j, \, (i,j)\notin\E, \\
      -P_{g_{ij}}, & i \neq j, \, (i,j)\in\E, \\ 
      \sum_{k\in\N_i}P_{g_{ik}}, & i=j, \, i\in\V, \\
  \end{array}
\right.
\end{align*}
where $[\L]_{ij}\in\R^{d\times d}$ is the $ij$th block of submatrix of $\L$. The bearing Laplacian can be viewed a matrix-weighted Laplacian which describes both the underlying graph and the inter-neighbor bearings of the network. See Figure~\ref{fig_demoBearingLaplacian} for illustration.

For undirected graphs, the bearing Laplacian has the following properties \cite[Lemma~2]{zhao2015NetLocalization}:
\begin{enumerate}[(a)]
     \item $\L$ is symmetric and positive semi-definite because for any $x=[x_1^T ,\dots,x_n^T ]^T \in\R^{dn}$
     $$x^T \L x=\frac{1}{2}\sum_{i\in\V}\sum_{j\in\N_i}(x_i-x_j)^TP_{g_{ij}}(x_i-x_j)\ge0.$$
     \item $\rank(\L)\le dn-d-1$ and $\myspan\{\one\otimes I_d, p\}\subseteq \Null(\L)$ for any network.
     \item $\rank(\L)=dn-d-1$ and $\Null(\L)=\myspan\{\one\otimes I_d, p\}$ if and only if the network is infinitesimally bearing rigid.
\end{enumerate}
In a network with $n_a$ anchors and $n_f=n-n_a$ followers, the bearing Laplacian may be partitioned into
\begin{align*}
    \L=\left[
         \begin{array}{cc}
           \L_{aa} & \L_{af} \\
           \L_{fa} & \L_{ff} \\
         \end{array}
       \right],
\end{align*}
where $\L_{ff}\in\R^{dn_f \times dn_f}$.
For any network, $\L_{ff}$ is positive semi-definite and satisfies $\L_{ff}p_f=-\L_{fa}p_a$ \cite[Lemma~3]{zhao2015NetLocalization}.
In the context of formation control, the anchors are called leaders and the subscript $a$ is replaced by $\ell$.

\newpage
\section*{Laman Graphs and Henneberg Construction}\label{sidebar_LamanGraph}

\begin{figure}[h]
  \centering
\includegraphics[width=\linewidth]{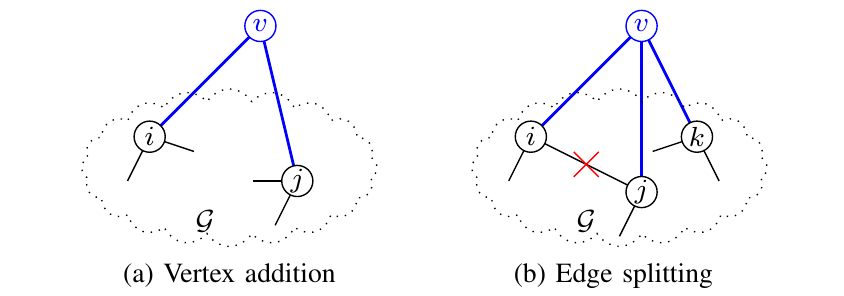}
  \caption{The two operations of the Henneberg construction.  The Henneberg construction can be used to generate all minimally infinitesimally distance rigid graphs in the plane.  The main idea is to ensure that the vertex addition and edge splitting operations satisfy the Laman condition at each step.}
  \label{fig_demoHennebergConstruction}
\end{figure}

An undirected graph $\G=(\V,\E)$ is called \emph{Laman} if $m=2n-3$ and every subset of $k\ge2$ vertices spans at most $2k-3$ edges \cite{Laman1970}.
Laman graphs can be characterized by the Henneberg construction as described below.
Given a graph $\G=(\V,\E)$, a new graph $\G'=(\V',\E')$ is formed by adding a new vertex $v$ to $\G$ and performing one of the following two operations:
\begin{enumerate}[(a)]
\item \emph{Vertex addition}: connect vertex $v$ to any two existing vertices $i,j\in\V$. In this case, $\V'=\V\cup \{v\}$ and $\E'=\E\cup\{(v,i),(v,j)\}$. See Figure~\ref{fig_demoHennebergConstruction}(a) for illustration.
\item \emph{Edge splitting}: consider three vertices $i,j,k\in\V$ with $(i,j)\in\E$ and connect vertex $v$ to $i,j,k$ and delete $(i,j)$. In this case, $\V'=\V\cup \{v\}$ and $\E'=\E\cup\{(v,i),(v,j),(v,k)\}\setminus\{(i,j)\}$. See Figure~\ref{fig_demoHennebergConstruction}(b) for illustration.
\end{enumerate}

A Henneberg construction starting from an edge connecting two vertices leads to a Laman graph \cite{WhiteleyHenneberg1985,Whiteley2005Pseudotriangulation,Jackson2007NotesRigidity}. The converse is also true. That is if a graph is Laman, then it can be generated by a Henneberg construction \cite[Lemma~2]{Whiteley2005Pseudotriangulation}.
The underlying graphs of the networks in Figure~\ref{fig_IPRExamples}(a)--(c) are Laman.
Laman graphs play critical roles in the construction of distance rigid and bearing rigid networks.

\newpage
\section*{Comparison of Bearing Rigidity and Distance Rigidity}\label{sidebar_compareBearingDistanceRigidity}

Both of the bearing and distance rigidity theories address the same problem of when the geometric pattern of a network can be uniquely determined. The difference is that the bearing rigidity theory considers inter-neighbor bearings whereas the distance rigidity theory considers inter-neighbor distances.
The term ``unique pattern'' in the bearing rigidity theory means the location of a network can be determined up to a translational and scaling factor, while in the distance rigidity theory it means the network can be determined up to a translational and rotational factor.

One connection between the two rigidity theories is that infinitesimal bearing rigidity is equivalent to infinitesimal distance rigidity in two dimensions \cite[Theorem~8]{zhao2014TACBearing}. In other words, a network in the plane is infinitesimally bearing rigid if and only if it is infinitesimally distance rigid. This equivalence property explains why the distance rigidity theory could be used to analyze the problems of bearing-based network localization or formation control in the literature \cite{Piovan2013Automatica,ZhuGuangwei2014Automatica,Eric2014ACC}.
It also suggests that the infinitesimal distance rigidity of a network by be examined by its infinitesimal bearing rigidity.
For example, it may not be straightforward to see that the networks in Figure~\ref{fig_nonIBRExamples}(c)-(d) are not infinitesimally distance rigid. However, it is intuitive to see these networks are not infinitesimally bearing rigid because there exist nontrivial infinitesimal bearing motions.
It must be noted that the equivalence cannot be generalized to three or higher dimensions. For example, the three-dimensional networks shown in Figure~\ref{fig_IPRExamples}(c)-(e) are infinitesimally bearing rigid but not infinitesimally distance rigid.

Compared to infinitesimal distance rigidity, infinitesimal bearing rigidity possess some interesting properties.
First, infinitesimal bearing rigidity not only ensures the unique pattern of a network, but also can be examined by a rank condition easily.
As a comparison, infinitesimal distance rigidity may not be able to ensure a unique pattern though it can be examined by a rank condition.
Second, an infinitesimally bearing rigid network remains infinitesimally bearing rigid when the dimension is lifted up to a higher dimension \cite[Theorem~7]{zhao2014TACBearing}. As a comparison, a network that is infinitesimally distance rigid in the plane may be flexible in a higher dimension.
Third, in the bearing rigidity theory, a Laman graph is generically bearing rigid in arbitrary dimensions and at most $2n-3$ edges would be sufficient to guarantee the bearing rigidity of a network in an arbitrary dimension. As a comparison, although a Laman graph embedded in a generic configuration is infinitesimally distance rigid \cite{WhiteleyHenneberg1985,Whiteley2005Pseudotriangulation,Jackson2007NotesRigidity,Anderson2008CSM,ConnellyBook}, this result, known as Laman's Theorem \cite{Laman1970}, is valid merely in two dimensional spaces. In three or higher dimensions, extra conditions and more edges are required to guarantee distance rigidity. The above comparison is summarized in Table~\ref{table_compareBearingDistanceRigidity}.

Why bearing rigidity has appealing properties in high dimensions can be explained intuitively from the perspective of degree of freedom. For example, consider a network of $n$ nodes in $d$-dimensional space. The network has $dn$ degrees of freedoms. In order to ensure the rigidity of the network, there must exist sufficient distance or bearing constraints to reduce the degrees of freedom of the network to certain desired values. Given a distance rigid network, when lifted up to a higher dimension, the degrees of freedom of the network increases while the number of constraints posed by an inter-neighbor distance remain the same. As a result, in order to preserve distance rigidity in higher dimensions, more distance constraints are required. As a comparison, when lifted to a higher dimension, the number of independent constraints posed by an inter-neighbor bearing also increases. For example, a bearing in the plane is equivalent to an azimuth angle whereas a bearing in the three dimensional space is equivalent to two bearing angles: azimuth and altitude. As a result, the same number of bearings are still able to preserve the bearing rigidity of the network.

\newpage
\begin{table*}  
      \caption{Comparison of infinitesimal bearing rigidity and infinitesimal distance rigidity.} \label{table_compareBearingDistanceRigidity}
      \centering
      \renewcommand{\arraystretch}{1.5} 
      \begin{tabular}{|l|p{6cm}|p{6cm}|}
        \hline
          & \textbf{Infinitesimal Bearing Rigidity (IBR)} &  \textbf{Infinitesimal Distance Rigidity (IDR)} \\
        \hline
        \textbf{Unique geometric pattern} & Yes, IBR ensures the unique pattern of a network. & No, IDR does not ensure the unique pattern of a network (global distance rigidity does).\\
        \hline
        \textbf{Rank condition} & Yes, IBR corresponds to a rank condition of the bearing rigidity matrix. & Yes, IDR corresponds to a rank condition of the distance rigidity matrix. \\
        \hline
        \textbf{Invariance to dimension} & Yes, a network that is IBR in a lower dimension remains IBR in a higher dimension. & No, a network that is IDR in a lower dimension may be flexible in a higher dimension. (Universal distance rigidity is invariant to dimensions)\\
        \hline
        \textbf{Minimum edge number} & In an arbitrary dimension, $2n-3$ edges are sufficient to ensure IBR. Less than $2n-3$ edges may also be sufficient to ensure IBR in three or higher dimensions. & In the plane, $2n-3$ is the minimum number of edges to ensure IDR. More than $2n-3$ edges are required to ensure IDR in three or higher dimensions.\\
        \hline
        \textbf{Laman graphs} &  In an arbitrary dimension, Laman graphs mapped to almost all configurations result in IBR networks. & In the plane, Laman graphs mapped to almost all configurations result in IDR networks. A similar result does not exist in higher dimensions. \\
        \hline
      \end{tabular}
\end{table*}

\newpage
\section*{Bearing Rigidity Theory for $SE(2)$}\label{sidebar_SE2Rigidity}

Consider a collection of $n$ nodes in $\mathbb{R}^2 \times \mathcal{S}^1$. Each point is described by its position $p_i \in \mathbb{R}^2$ and its orientation $\psi_i \in \mathcal{S}^1$.  An $SE(2)$ network, denoted as $(\G,p,\psi)$, is the directed graph $\mathcal{G}=(\mathcal{V},\mathcal{E})$, and the configuration $(p,\psi)$, where each vertex $i \in \mathcal{V}$ in the graph is mapped to the point $(p_i,\psi_i) \in SE(2)$.  Note that $SE(2)$ networks, \emph{directed} graphs are considered.

\begin{figure}
  \centering
\includegraphics[width=\linewidth]{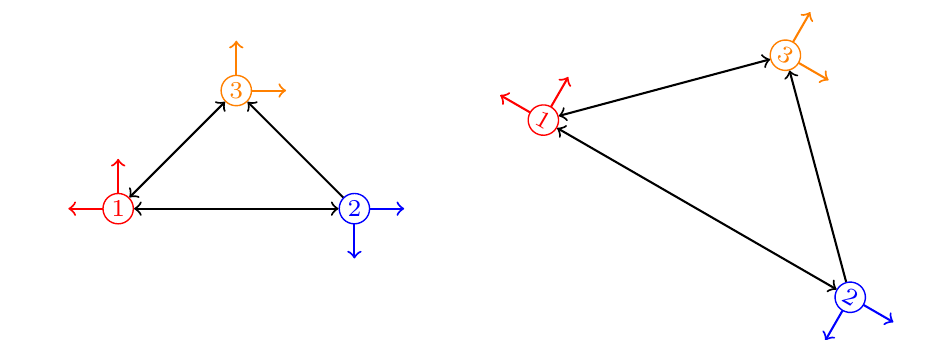}
  \caption{Example of two congruent $SE(2)$ networks.  The above two networks differ in terms of a translation, a scaling, and a coordinated rotation.}
  \label{fig_demoSE2RigidityConcept}
\end{figure}

Suppose $(i,j) \in \mathcal{E}$ is the $k$th directed edge where $k=\{1,\dots,m\}$ and $m$ denotes the number of directed edges in $\E$. Let $g_k$ be the relative bearing of $p_j$ with respect to $p_i$ expressed in the global frame. Then,
$$r_k = \left[
         \begin{array}{cc}
           \cos\psi_i & \sin\psi_i \\
           -\sin\psi_i & \cos\psi_i \\
         \end{array}
       \right]g_k$$
is the bearing $g_k$ expressed in node $i$'s local reference frame.
Define the \emph{directed bearing function} associated with the $SE(2)$ network, $F_{SE} : SE(2)^{n} \rightarrow {\mathcal{S}^2}^{m}$, as
\begin{align}\label{bearing_rigidity_fcn}
F_{SE}(p,\psi) &= [ r_{1}^T \; \cdots \; r_{m}^T ]^T \in {\mathcal{S}^2}^{m}.
\end{align}
The corresponding \emph{directed bearing rigidity matrix} is defined as the Jacobian of the directed bearing function,
\begin{align}\label{eq_rigidityMatrixDefinition}
    R_{SE}(p,\psi) \triangleq \frac{\partial F_{SE}(p,\psi)}{\partial (p,\psi)}\in\R^{2m\times 3n}.
\end{align}
Let $\delta \chi\in\R^{3n}$ be a variation of the configuration $(p,\psi)$.  If $R_{SE}(p,\psi) \delta \chi = 0$, then $\delta \chi$ is called an \emph{infinitesimal $SE(2)$ bearing motion} of $\mathcal{G}(p,\psi)$. There are three types of trivial infinitesimal $SE(2)$ motions corresponding to \emph{translations}, \emph{scalings}, and \emph{coordinated rotations} of the entire network. The coordinated rotation involves an angular rotation of each agent about its own body axis with a rigid-body rotation of the network (see Figure \ref{fig_demoSE2RigidityConcept}).
An $SE(2)$ network is \emph{infinitesimally bearing rigid} if all the infinitesimal bearing motions are trivial. A necessary and sufficient condition for an $SE(2)$ network to be infinitesimally bearing rigid is \cite{zelazo2015CDC,Schiano2016}
$$\rank [R_{SE}(p,\psi)] = 3n - 4,$$
or equivalently
$$\Null [R_{SE}(p,\psi)] =  \myspan \left\{ \left[
         \begin{array}{c} \one_n \otimes I_2 \\ 0 \end{array}\right],  \left[
         \begin{array}{c} p \\ 0 \end{array}\right],  \left[
         \begin{array}{c} p^\perp \\ \one_n \end{array}\right] \right\},$$
where $p^\perp=[(p_1^\perp)^T,\dots,(p_n^\perp)^T]^T$ and $p_i^\perp=R_{\pi/2}p_i$. The null-space is characterized in this way after a permutation of the matrix that groups the positions and attitudes of all agents together. Here $R_{\pi/2}$ is a rotation matrix that rotates any vector by $\pi/2$.

Detailed definitions in the $SE(2)$ bearing rigidity theory can be found in  \cite{zelazo2014SE2Rigidity,zelazo2015CDC,Schiano2016}. The $SE(2)$ rigidity theory has been employed for distributed relative position estimation \cite{zelazo2014SE2Rigidity} and formation control \cite{zelazo2015CDC, Schiano2016, Schiano2017}.  A similar approach has been extended for $SE(3)$ \cite{Michieletto2016CDC}.

\clearpage

\newpage
\section*{Author Information}

Shiyu~Zhao is a Lecturer in the Department of Automatic Control and Systems Engineering at the University of Sheffield, UK. He received the B.E. and M.E. degrees from Beijing University of Aeronautics and Astronautics, China, in 2006 and 2009, respectively. He obtained the Ph.D. degree in Electrical Engineering from the National University of Singapore in 2014. From 2014 to 2016, he served as post-doctoral researchers at the Technion - Israel Institute of Technology and University of California at Riverside. He is a corecipient of the Best Paper Award (Guan Zhao-Zhi Award) in the 33rd Chinese Control Conference, Nanjing, China, in 2014. His research interests lie in networked dynamical systems and unmanned aerial vehicles.

Daniel~Zelazo is an Assistant Professor of Aerospace Engineering at the Technion - Israel Institute of Technology, Israel. He received the B.S. (99) and M.Eng. (01) degrees in Electrical Engineering from the Massachusetts Institute of Technology. In 2009, he completed his Ph.D. from the University of Washington in Aeronautics and Astronautics. From 2010 to 2012, he served as a post-doctoral research associate and lecturer at the Institute for Systems Theory \& Automatic Control in the University of Stuttgart. His research interests include topics related to multi-agent systems, optimization, and graph theory.

\end{document}